\documentclass[vecphys]{svmult}

\usepackage{graphicx}        % standard LaTeX graphics tool
\usepackage[bottom]{footmisc}% places footnotes at page bottom
\begin{document}

\title*{Spectroscopic Properties of Polarons in Strongly Correlated 
Systems by Exact Diagrammatic Monte Carlo Method}
\titlerunning{Spectroscopic Properties of Polarons by Exact Monte Carlo} 
% for an abbreviated version of
\author{A. S. Mishchenko\inst{1,2}\and
N. Nagaosa\inst{3}}
\institute{CREST, Japan Science and Technology Agency (JST),
AIST, 1-1-1, Higashi, Tsukuba 305-8562, Japan.
\and Russian Research Centre ``Kurchatov Institute'', 
123182 Moscow, Russia.
\and Department of Applied Physics, The University of Tokyo, 7-3-1
Hongo, Bunkyo-ku, Tokyo 113, Japan.}
\maketitle

\section{Introduction}
\label{section_1}

Theoretical study of polarons in the strongly correlated system is like an 
attempt to view contents of a Pandora box embedded into another, even more 
sinister and obscure, container of riddles, enigmas and mysteries. 
This desperate situation occurs because solution is not known even for the 
simplest polaron problem, i.e.\ when a perfectly stable quasiparticle (QP) 
with momentum as a single quantum number interacts with a well defined bath 
of bosonic elementary excitations. 
To the contrary, the definition of the strongly correlated system implies 
that QPs might be highly unstable and the very notion of QPs, both in 
electronic and bosonic subsystems, is under question. 
Thus, one faces the problem of an interplay between ill defined objects and 
it is crucial to solve the problem without approximations.  
Further difficulty, pertinent to realistic systems, is an interplay of the 
momentum and other quantum numbers characterizing internal states of a QP. 
       
The problem of polaron originally emerged as that of an electron coupled to 
phonons (see \cite{Appel,Pekar51}).
In the initial formulation a structureless QP is characterized by the only 
quantum number, momentum, which changes due to interaction of the QP with 
phonons \cite{Landau33,Frohlich}. 
Later, depending on what can be called ``particle'' and ``environment'', 
and how they interact with each other, the polaron concept was related 
to extreme diversity of physical phenomena. 
There are many other objects which, having nothing to do with phonons, 
are isomorphic to simple polaron \cite{DevEnciclop}, as, e.g. an 
exciton-polaron in the intraband scattering approximation 
\cite{Ansel,UKKTH86,Toyo,Toyobook}.
Another example is the problem of a hole in the antiferromagnet which is 
closely related to polaron since hole movement is accompanied by the spin 
flips which, in the spin wave approximation, are equivalent to creation and 
annihilation of magnons \cite{Kane,Izymov97}. 

The concept of polaron was further generalized to include internal degrees of 
freedom which, interacting with environment, change their quantum numbers. 
Example of a complex QP is the Jahn-Teller polaron, where electron-phonon 
interaction (EPI) changes quantum numbers of degenerate electronic states 
\cite{Kanamori,Abragam,Kugel}. 
This generalization is important due to it's relevance to the colossal 
magnetoresistance phenomena in the manganese oxides 
\cite{Millis95,AlexBrat99}. 
Another example is the pseudo Jahn-Teller polaron, where EPI is inelastic 
and leads to transitions between close in energy electronic levels of a QP     
\cite{RashPJT,ToyPJT,BersuBook}. 
Further generalization is a system of several QPs which interact both with 
each other and environment.
For example, effective interaction of two electrons through exchange by 
phonons can overcome the Coulomb repulsion and form a bound state, bipolaron  
\cite{Vinetskii,AnderBP,HiramToy,AlexRunn,AlexRun2}. 
On the other hand, coupling of attracting hole and electron to the 
lattice vibrations \cite{Haken56,Bassani75,Pollman77}
can create a lot of qualitatively different objects: 
localized exciton, weakly bound pair of localized hole and localized electron, 
etc.\ \cite{Sumi77,UKKTH86}.
Scattering by impurities introduces additional complexity to the polaron 
problem because interference of impurity potential with lattice distortion, 
which accompanies the polaron movement, can contribute either constructively 
or destructively to the localization of a QP on impurity 
\cite{Primesi1,Primesi2,UKKTH86}. 

In addition, a bare QP and bosonic bath can not be considered as well defined 
in the correlated systems. 
Angle Resolved Photoemission Spectra (ARPES), revealing the Lehmann
Function (LF) of quasiparticle, demonstrate broad peaks in many 
correlated systems: cooper oxide high-temperature superconductors 
\cite{ZX95,Shen_03,DichotomyZhou}, colossal magnetoresistive manganites 
\cite{Dessau_98,Manella_04,Manella_05},
quasi-one-dimensional Peierls conductors \cite{Perf_01, Perf_02},
and Verwey magnetites \cite{Schrupp_05}. 
Besides, phonons are also broadened in many correlated systems, e.g.\ in 
high-temperature semiconductors \cite{Pintsch} and mixed-valent materials 
\cite{Mook79,Mook82}. 
One of possible reasons for these broadenings is the interaction of the QPs 
with the lattice degrees of freedom. 
However, in many realistic cases other subsystems, not explicitly included 
into the polaron Hamiltonian, are responsible for the decay of QP and phonons, 
e.g., another electronic bands, phonon anharmonicity, interaction with 
nuclear spins, etc. 
Then, if this auxiliary broadening is known in some approximation, one can 
formulate an ambitious goal to study spectral response when ``bare'' 
quasiparticle with known damping interacts with ``broadened'' 
bosonic excitations.

No one of traditional numerical methods, to say nothing of analytical ones, 
can give \textit{approximation free 
results for measurable quantities} of polaron, such as optical conductivity
or angle resolved photoemission spectra, for \textit{in macroscopic 
system of arbitrary dimension}.  
Besides, we are not aware of any numerical method which can  incorporate in 
an approximation free way the information on the damping of QP and bosonic 
bath. 
Below we describe basics of recently developed Diagrammatic Monte Carlo 
(DMC) method for numerically exact computation of Green functions and 
correlation functions in imaginary time for few polarons in a macroscopic 
system 
\cite{Zhetf,PS,MPSS,Exciton,Aus02,Optics,tJpho,UFN_05,JPSJ_exc}. 
Analytic continuation of imaginary time functions to real frequencies 
is performed by a novel approximation free approach of stochastic optimization
(SO) \cite{MPSS,UFN_05,JPSJ_exc}, circumventing difficulties of popular 
Maximal Enthropy method.
Finally we focus on results of 
application of the DMC-SO machinery to various problems
\cite{MN01,Hole01,RP02,Isotope,tJphoSpain,FrCondon} 

The basic models, related to the polaronic objects in correlated systems, 
which can be solved by DMC-SO methods, are stated in the
next Sect. 
It is followed in Sect.\ \ref{sec_1_subsec_2} by description of 
stumbling blocks encountered by analytic methods. 
Sect.\ \ref{section_2} concerns the basics of DMC-SO methods. 
However, those who are not interested in the details of the methods can 
briefly look through the definitions in the introduction for Sect.\ 
\ref{section_2} and turn to Sect.\ \ref{section_3} where LF and optical 
conductivity of Fr\"{o}hlich  polaron are discussed
(see also \cite{D}). 
Results of studies of the self-trapping phenomenon are presented in Sect.\ 
\ref{section_4} and application of DMC-SO methods to the exciton problem 
can be found in Sect.\ \ref{section_5}. The chapter is completed by 
Sect.\ \ref{section_6} devoted to studies of ARPES of high temperature 
superconductors.

\subsection{Formulation of a General Model with Interacting Polarons}
\label{sec_1_subsec_1}

In general terms, the simplest problem of a complex polaronic object, 
where center-of-mass motion does not separate from the rest of degrees of 
freedom, is introduced as system of two QPs
\begin{equation}
\hat{H}_{0}^{\mbox{\scriptsize par}}
=\sum_{{{\bf k}}}\varepsilon _{a}({\bf k})a_{{
{\bf k}}}^{\dagger }a_{{ {\bf k}}}+\sum_{{\bf k}}\varepsilon _{h}(%
{\bf k})h_{{ {\bf k}}}h_{{ {\bf
k}}}^{\dagger }  \label{1}
\end{equation}
($a_{{ {\bf k}}}$ and $h_{{ {\bf k}}}$ are annihilation operators, and  
$\varepsilon _{a}({\bf k})$ and $\varepsilon _{h}({\bf k})$ 
are dispersions of QPs), which interact with each other  
\begin{equation}
\hat{H}_{\mbox{\scriptsize a-h}}=-N^{-1}\sum_{{{\bf pkk^{\prime }}}}%
{\cal U}({\bf p},{\bf k},{\bf k}^{\prime })a_{{{\bf p+k}}%
}^{\dagger }h_{{ {\bf p-k}}}^{\dagger }h_{{ {\bf %
p-k^{\prime }}}}a_{{ {\bf p+k^{\prime }}}}.  \label{2}
\end{equation}
($N$ is the number of lattice sites) through the instantaneous 
Coulomb potential and the scattering by bosons
\begin{eqnarray}
&\hat{H}_{\mbox{\scriptsize par-bos}} =
i \sum\limits_{\kappa=1}^{Q} 
\sum\limits_{{\bf k},{\bf q}} 
(b^{\dagger}_{{\bf q},\kappa} - b_{-{\bf q},\kappa})& \nonumber \\
&\left[
\gamma_{aa,\kappa}({\bf k},{\bf q}) a_{{\bf k}-{\bf q}}^{\dagger} a_{{\bf k}} +
\gamma_{hh,\kappa}({\bf k},{\bf q}) h_{{\bf k}-{\bf q}}^{\dagger} h_{{\bf k}} +
\gamma_{ah,\kappa}({\bf k},{\bf q}) h_{{\bf k}-{\bf q}}^{\dagger} a_{{\bf k}} 
\right]
+ h.c. &
\label{3}
\end{eqnarray}
($\gamma_{[aa,ah,hh],\kappa}$ are interaction constants) where   
quanta of Q different branches of bosonic excitations are created or
annihilated, which are described by
\begin{equation}
\hat{H}_{\mbox{\scriptsize bos}} =
\sum_{\kappa=1}^{Q} 
\sum_{\bf q} \omega_{{\bf q},\kappa} 
b_{{\bf q},\kappa}^{\dagger} b_{{\bf q},\kappa} \; .
\label{4}
\end{equation}
In general, each QP can be a composite one with internal degree of freedom
represented by $T$ different states 
\begin{equation}
\hat{H}_0^{\mbox{\scriptsize PJT}} =
\sum_{{\bf k}} \sum_{i=1}^{T} \epsilon_{i}({\bf k})
a_{i,{\bf k}}^{\dagger} a_{i,{\bf k}},
\label{5}
\end{equation}
which quantum numbers can be also changed due to nondiagonal part 
of particle-boson interaction 
\begin{equation}
\hat{H}_{\mbox{\scriptsize par-bos}} =
i \sum_{\kappa}^{Q} \sum_{{\bf k},{\bf q}} \sum_{i,j=1}^{T}
\gamma_{ij,\kappa}({\bf k},{\bf q}) (b^{\dagger}_{{\bf q},\kappa} - b_{-{\bf 
q},\kappa})
a_{i,{\bf k}-{\bf q}}^{\dagger} a_{j,{\bf k}} + h.c.
\label{6}
\end{equation}   
Complicated model (\ref{1})-(\ref{6}) is still too far from the cases 
encountered in strongly correlated systems.
Due to coupling of QPs (\ref{1}) and (\ref{5}) and bosonic fields (\ref{4}) 
to additional degrees of freedom, these excitations are not well defined 
from the onset. 
Namely, the dispersion relation of the QP spectrum $\epsilon({\bf k})$ in 
realistic system is ill-defined. One can speak of a Lehmann Function (LF) 
\cite{AGD,Mahan,JG} of a QP
\begin{equation}
L_{\bf k} (\omega) \, = \, \sum_{\nu} \,
\delta(\omega - E_{\nu}({\bf k})) \;
\vert \langle \nu \vert a^{\dag}_{\bf k} \vert \mbox{vac} \rangle \vert^2 
\; 
\label{g}
\end{equation}
,which is normalized to unity 
%\begin{equation}
$ \int_{0}^{+\infty} d \omega L_{\bf k} (\omega) = 1 $
%\label{g_norm}
%\end{equation}
and can be interpreted as a probability that a QP has momentum ${\bf k}$ and
energy $\omega$. 
(Here $\{ \vert \nu \rangle \}$ is a complete set of eigenstates of 
Hamiltonian $\hat{H}$ in a sector of given momentum ${\bf k}$:  
$H \, \vert \nu ({\bf k}) \rangle = 
E_{\nu}({\bf k}) \, \vert \nu ({\bf k}) \rangle$.)
Only for noninteracting system the LF reduces 
to delta function
%\begin{equation}
$L_{\bf k}^{\mbox{\scriptsize NONINT}} (\omega) \, = \, 
\delta (\omega - \epsilon({\bf k})) $
%\label{g_nonint}
%\end{equation}
and, thus, sets up dispersion relation $\omega=\epsilon({\bf k})$.

Specific cases of model (\ref{1})-(\ref{6}) describe enormous variety of 
physical problems. Hamiltonians (\ref{1}) and (\ref{2}), in case of attractive 
potential  ${\cal U}({\bf p},{\bf k},{\bf k}^{\prime })>0$, describe an 
exciton with static screening \cite{Knox,Egri}. 
Besides, expressions (\ref{1})-(\ref{4}) describe bipolaron 
for repulsive interaction \cite{Vinetskii,AnderBP,HiramToy,AlexRunn,AlexRun2}
${\cal U}({\bf p},{\bf k},{\bf k}^{\prime })<0$ and exciton-polaron 
otherwise \cite{Haken56,Bassani75,Pollman77}.
The simplest model for exciton-phonon interaction, when only two ($T=2$) 
lowest states of relative electron-hole motion are relevant 
(e.g. in one-dimensional charge-transfer exciton 
\cite{Haarer75,HaPhi75,ElsWeis85}), 
is defined by Hamiltonians (\ref{4})-(\ref{6})). 
The same relations (\ref{4})-(\ref{6}) describe the problems of Jahn-Teller  
[all $\epsilon_{i}$ in Hamiltonian (\ref{5}) are the same] and pseudo 
Jahn-Teller polaron. 
The problem of a hole in an antiferromagnet in spin-wave approximation 
is expressed in terms of Hamiltonians (\ref{4})-(\ref{6}) with $Q=1$ and 
$T=1$. 
When hole also interacts with phonons, one has to take into account one more 
bosonic branch and set $Q=2$ in (\ref{4}) and (\ref{6}).
Finally, the simplest nontrivial problem of a polaron, i.e.\ of a 
structureless QP  interacting with one phonon branch, is described 
by noninteracting Hamiltonians of QP $\hat{H}_{\mbox{\scriptsize par}}$ 
and phonons $\hat{H}_{\mbox{\scriptsize ph}}$ 
\begin{equation}
 \hat{H}_{\mbox{\scriptsize 0}} =
\sum_{{\bf k}} \epsilon({\bf k}) a_{{\bf k}}^{\dagger} a_{{\bf k}} 
+
\sum_{\bf q} \omega_{{\bf q}} 
b_{{\bf q}}^{\dagger} b_{{\bf q}} \; ,
\label{7}
\end{equation} 
and interaction term
\begin{equation}
\hat{H}_{\mbox{\scriptsize int}} =
\sum_{{\bf k},{\bf q}} 
V({\bf k},{\bf q}) (b^{\dagger}_{{\bf q}} - b_{-{\bf q}})
a_{{\bf k}-{\bf q}}^{\dagger} a_{{\bf k}} + h.c.
\label{9} \; .
\end{equation}
The simplest polaron problem, in turn, can be subdivided into continuous
and lattice polaron models.

\subsection{Limitations of Analytic Methods in Problem of Polarons}
\label{sec_1_subsec_2}

Analytic solution for the problem of exciton in a rigid lattice
is available only for small radius Frenkel regime \cite{Frenkel} 
and large radius Wannier regime \cite{Wannier}. 
However, even limits of validity for these approximations are not known. 
Random phase approximation approaches \cite{Knox,Egri}, are capable of 
obtaining some qualitative conclusions for intermediate radius regime though 
its' quantitative results are not reliable due to uncontrolled errors. 
The situation is similar with the problem of structureless polaron, where 
analytic solutions are known only in the weak and strong coupling regimes. 
Besides, reliable results for these regimes are available only for ground 
state properties. 

Although several novel methods, capable of obtaining properties of excited 
states, were developed recently, variational coherent-states expansion 
\cite{DeFilCat05} and free propagator momentum average summation 
\cite{Berciu} as a few examples, all of them, to provide reliable data in 
a specific regime, need either comparison with exact sum rules 
\cite{DevreLeRo76,KornilEL02} or with exact numerical results. 

Application of variational methods to study of excitations is a tricky 
issue since, strictly speaking, they are valid only for the ground state.
As an example for the importance of sum rules in variational treatment, we 
refer to the problem of the optical conductivity of the Fr\"{o}chlich polaron. 
Possibility of existence of Relaxed Excited State (RES), which is a metastable 
state where lattice deformation has adjusted to the electronic 
excitation rendering stability and narrow linewidth of the spectroscopic 
response, was briefly mentioned by S.\ I.\ Pekar in early 50's. Then,
conception of RES was rigorously formulated by J. T. Devreese with coworkers
and has been a subject of extensive investigations for years 
\cite{DevEnciclop,Dev66,KED69,Dev72,DeRev72,Dev_twopho_73,Optics,FrCondon}. 
Calculations of impedance \cite{Dev72} in the framework of technique 
\cite{Fey62} supported the existence of a narrow stable peak in the 
optical conductivity.
However, even the authors of \cite{Dev72} were skeptical about the fact 
that the width of RES in the strong coupling regime appeared to be more 
narrow than the phonon frequency, i.e. inverse time which is, according 
to the Heisenberg uncertainty principle, is required for the lattice 
readaptation. In consequent paper \cite{Dev_twopho_73} they realized the 
importance of many-phonon processes and studied two-phonon contribution to 
optical conductivity. 
Importance of many phonon processes was confirmed when variational results 
\cite{Dev72} were compared with exact DMC simulations \cite{Optics}. 
Variational result well reproduced the position of the 
peak in exact data though failed in description of the peak width in the 
strong coupling regime \cite{Optics}. Finally, when approach \cite{Dev72}
was modified and several sum rules were accurately introduced into 
variational model \cite{FrCondon}, both position and width of the peak were 
quantitatively reproduced. Studies \cite{FrCondon} (see 
Sect.\ \ref{sec_3_subsec_1}), do not address rather philosophical question 
whether RES exists or not, though inevitably prove that, in contrast 
to the foregoing beliefs, there in no stable excited state of the 
Fr\"{o}hlich polaron in the strong coupling regime.    
Note that sometimes excited states can not be handled by analytic methods 
even for weak couplings: perturbation theory expression for LF of the 
Fr\"{o}hlich polaron model diverges at the phonon energy 
$\omega_{\mbox{\scriptsize ph}}$  [See (\ref{spa_2}) in Sect.\ 
\ref{sec_3_subsec_1}.] and more elaborate treatment is necessary. 

Difficulties of semianalytic methods enhance in the intermediate coupling 
regime where results are sometimes wrong even for ground state properties. 
For example, 
the variatioanl approach \cite{LLP53}, which has been considered as an 
intermediate coupling theory, appeared to be valid only in the weak coupling 
limit \cite{MPSS}.   
Special interest to the methods, giving reliable information on
excited states, is triggered by the self-trapping phenomenon which occurs 
just in the intermediate coupling regime. 
This phenomenon is a dramatic transformation of QP properties when system 
parameters are slightly changed \cite{Landau33,UKKTH86,Toyobook,R82}.
In the intermediate coupling regime ``trapped'' QP state with strong lattice 
deformation around it and ``free'' state with weakly perturbed lattice may 
hybridize and resonate because of close energies at some critical value of 
electron-lattice interaction $\gamma_c$. 
It is clear that, to study self-trapping, one has to apply a method 
giving reliable information on excited states in the intermediate coupling 
regime.

\section{Diagrammatic Monte Carlo and Stochastic Optimization Methods}
\label{section_2}

In this section we introduce definitions of exciton-polaron properties which 
can be evaluated by DMC and SO methods. 
An idea of DMC approach for numerically exact calculation of Green functions 
(GFs) in imaginary times is presented in Sect.\ \ref{sec_2_subsec_1}, and
a short description of SO method, which is capable of making unbiased analytic 
continuation from imaginary times to real frequencies, is given in 
Sect.\ \ref{sec_2_subsec_2}.
Using combination of DMC and SO, one can often circumvent difficulties of 
analytic and traditional numerical methods. 
Therefore, a brief comparative analysis of advantages and drawbacks of DMC-SO 
machinery is given in Sect.\ \ref{sec_2_subsec_3}. 

To obtain information on QPs it is necessary to calculate Matsubara GF in 
imaginary time representation and make analytic continuation 
to the real frequencies \cite{Mahan}.
For the two-particle problem (\ref{1})-(\ref{4}) the relevant quantity is 
the two-particle GF \cite{Exciton,Aus02} 
\begin{equation}
G_{{\bf k}}^{{\bf pp}'}(\tau )=\langle
\mbox{vac}\mid a_{{ {\bf k+p^{\prime }}}}(\tau )h_{{ {\bf %
k-p^{\prime }}}}(\tau )h_{{ {\bf k-p}}}^{\dagger }a_{{
{\bf k+p}}}^{\dagger }\mid \mbox{vac}\rangle \; .
\label{10}
\end{equation}
(Here $h_{{\bf k-p}}(\tau )=e^{\hat{H}\tau }h_{{\bf k-p}}
e^{-\hat{H}\tau }$, $\tau >0$.) In the case of exciton-polaron, 
vacuum state $\mid \mbox{vac}\rangle$ is the state with filled 
valence and empty conduction bands. For the bipolaron problem it is a
system without particles. In the simpler case of a QP 
with two-level internal structure described by (\ref{4})-(\ref{6}) the 
relevant quantity is the one-particle matrix GF    
\cite{MN01,Aus02}
\begin{equation}
G_{{\bf k},ij}(\tau)=
\langle \mbox{vac} \mid
a_{i,{\bf k}}(\tau) a_{j,{\bf k}}^{\dagger }\mid \mbox{vac} \rangle,
\;\; i,j=1,2.
\label{11}
\end{equation}  
For a structureless polaron the matrix 
(\ref{11}) reduces to one-particle scalar GF
\begin{equation}
G_{{\bf k}}(\tau)=\langle \mbox{vac} \mid
a_{{\bf k}}(\tau) a_{{\bf k}}^{\dagger }\mid \mbox{vac} \rangle \; .
\label{12}
\end{equation} 
Information on the response to an external weak perturbation 
(e.g.\ optical absorption) is contained
in the current-current correlation function  
\mbox{$\langle J_{\beta} (\tau) J_{\delta} \rangle$}  
($\beta$/$\delta$ are Cartesian indexes).

Lehmann spectral representation of $G_{{\bf k}}(\tau)$ \cite{Mahan,JG} at 
zero temperature
\begin{equation}
G_{{\bf k}}(\tau) \, = \, \int_0^{\infty} d \omega \, L_{\bf k} (\omega) \,
e^{- \omega \tau} \; ,
\label{Fr}
\end{equation}
with the Lehmann function (LF) $L_{\bf k} (\omega)$ given in (\ref{g}),
reveals information on the ground and excited states.
Here $\{ \vert \nu \rangle \}$ is a complete set of eigenstates of 
Hamiltonian $\hat{H}$ in a sector of given momentum ${\bf k}$:  
$H \, \vert \nu ({\bf k}) \rangle = 
E_{\nu}({\bf k}) \, \vert \nu ({\bf k}) \rangle$.
The LF $L_{\bf k} (\omega)$ has poles 
(sharp peaks) on the energies of stable (metastable) states of particle. 
For example, if there is a stable state at energy $E({\bf k})$, 
the LF reads
%\begin{equation}
$L_{\bf k} (\omega) = Z^{({\bf k})} \, 
\delta(\omega - E({\bf k})) \, + \, \ldots$,
%\label{assym}
%\end{equation}
and the state with the lowest energy $E_{\mbox{\scriptsize g.s.}}({\bf k})$
in a sector of a given momentum ${\bf k}$ is highlighted by 
asymptotic behavior of GF
\begin{equation} 
G_{{\bf k}}(\tau \gg \max \left[ \omega_{{\bf q},\kappa}^{-1} \right] ) \; \to 
\; Z^{({\bf k} )} \, 
\exp [ -E_{\mbox{\scriptsize g.s.}}({\bf k}) \tau ] \; ,
\label{tends}
\end{equation}
where $Z^{({\bf k})}$-factor is the weight of the state. 
Analyzing the asymptotic behavior of similar 
$n$-phonon GFs \cite{MPSS,MN01} 
\begin{equation}
\begin{array}{l}  
\; \; \; \; \; \; \; \; \; \; \; \; \; \; \; \; \; \; \; \; \; \; \; \; \; \; \; 
G_{{\bf k}}(n, \tau ; \, {\bf q}_1, \ldots , {\bf q}_n ) \, = \, 
 \\
\langle \mbox{vac} \vert \,
b^{ }_{{\bf q}_n}({\tau}) \cdots b^{ }_{{\bf q}_1}({\tau})\,
a^{ }_{\bf p}(\tau ) a^{\dag}_{\bf p} \,
b^{\dag}_{{\bf q}_1} \cdots b^{\dag}_{{\bf q}_n}
\vert \mbox{vac} \rangle \; , \; \; 
{\bf p} \, = \, {\bf k} - \sum_{j=1}^n {\bf q}_j \; . 
\label{G_N} 
\end{array}
\end{equation}
one obtains detailed information about lowest state. For example, 
important characteristics of the lowest state wave function 
\begin{equation}
\Psi_{\mbox{\scriptsize g.s.}}({\bf k}) =  
\sum_{i=1}^{T} \sum_{n=0}^{\infty} \sum_{{\bf q}_1 ... {\bf q}_n}
\theta_i({\bf k}; {\bf q}_1,..., {\bf q}_n) c^{\dagger}_{i,{\bf k}-{\bf q}_1...-
{\bf q}_n}
b^{\dagger}_{{\bf q}_1} ... b^{\dagger}_{{\bf q}_n} \mid \mbox{vac} \rangle \; 
\label{vawefu}
\end{equation}
are partial $n$-phonon contribution
\begin{equation}
Z^{({\bf k})} (n) \equiv \sum_{i=1}^{T} \sum_{{\bf q}_1 ... {\bf q}_n} 
\mid
\theta_i({\bf k}; {\bf q}_1,..., {\bf q}_n)
\mid^2 \; 
\label{Zfactor}
\end{equation}
which is normalized to unity
\mbox{$\sum_{n=0}^{\infty} Z^{({\bf k})}(n) \equiv 1$}, and the average number
of phonons
\begin{equation}
\langle N \rangle \equiv 
\langle \Psi_{\mbox{\scriptsize g.s.}}({\bf k}) \mid \sum_{\bf q} 
b^{\dagger}_{\bf q} b_{\bf q} 
\mid \Psi_{\mbox{\scriptsize g.s.}}({\bf k}) \rangle =
\sum_{n=1}^{\infty} n Z^{({\bf k})} (n) \; 
\label{srefon}
\end{equation} 
in polaronic cloud. 
Another example is the wave function of relative electron-hole motion of 
exciton in the lowest state in the sector of given momentum
\begin{equation}
\Psi_{\mbox{\scriptsize g.s.}}({\bf k}) =
\sum_{\mbox{\scriptsize {\bf p}}} 
\xi_{\mbox{\scriptsize {\bf k p}}}(g.s.) 
a^{\dagger}_{\mbox{\scriptsize {\bf k+p}}}
h^{\dagger}_{\mbox{\scriptsize {\bf k-p}}} \mid \mbox{vac} \rangle \; .  
\label{WF}
\end{equation}
The amplitudes $\xi_{\mbox{\scriptsize {\bf k p}}}(g.s.)$ of this wave 
function can be obtained \cite{Exciton} from asymptotic behavior of the 
following GF (\ref{10})
\begin{equation}
G_{{\bf k}}^{{\bf  p=p}^{\prime}} (\tau \to \infty ) = 
\mid 
\xi_{\mbox{\scriptsize {\bf k p}}}(g.s.) \mid^2 
e^{-E_{\mbox{\scriptsize g.s.}}({\bf k}) \tau } \; . 
\label{examp}
\end{equation}

Information on the excited states is obtained by the analytic continuation
of imaginary time GF to real frequencies which requires to solve
the Fredholm equation
$G_{{\bf k}}(\tau) = \hat{\cal F} \left[ L_{\bf k} (\omega) \right]$ 
(\ref{Fr})
\begin{equation}
L_{\bf k} (\omega) = \hat{\cal F}^{-1}_{\omega} 
\left[ G_{{\bf k}}(\tau) \right] \; .
\label{Fredho}
\end{equation} 
The equation (\ref{Fr}) is a rather general relation 
between imaginary time GF/cor\-re\-la\-tor and spectral properties of the system.  
For example, the absorption coefficient of light by excitons 
${\cal I}(\omega)$ is obtained as solution of the same equation
\cite{Exciton}  
\begin{equation}
{\cal I}(\omega) = \hat{\cal F}^{-1}_{\omega} 
\left[\sum_{{\bf pp}'} G_{{\bf k}=0}^{{\bf pp}'}(\tau) \right] \; .
\label{Poglo}
\end{equation} 
Besides, the real part of the optical conductivity 
$\sigma_{\beta \delta} (\omega)$ is expressed \cite{Optics} in terms of 
current-current correlation function  
$\langle J_{\beta}(\tau ) J_{\delta} \rangle$ 
by relation
\begin{equation}
\sigma_{\beta \delta}(\omega) = 
\pi \hat{\cal F}^{-1}_{\omega} 
\left[ \langle J_{\beta}(\tau ) J_{\delta} \rangle \right] / \omega \; .
\label{Provodim}
\end{equation}

\subsection{Diagrammatic Monte Carlo Method}
\label{sec_2_subsec_1}
   
DMC Method is an algorithm which calculates GF (\ref{10})-(\ref{12}) 
without any systematic errors. This algorithm is described below for the 
simplest case of structureless polaron \cite{MPSS}, and generalizations to 
more complex cases can be found in consequent 
references\footnote{Generalization of described below technique to the case 
of exciton (\ref{1}-\ref{2}) is given in \cite{Exciton} and its modification 
for pseudo-Jahn-Teller polaron (\ref{4}-\ref{6}) is developed 
in \cite{MN01,Aus02}. Method for evaluation of current-current 
correlation function can be found in \cite{Optics} and a case of
a polaron interacting with two kinds of bosonic fields is considered 
in \cite{tJpho}.}.  
DMC is based on the Feynman expansion of the Matsubara GF in 
imaginary time in the interaction representation
\begin{equation}
G_{{\bf k}}(\tau) =
\left\langle \mbox{vac} \left\vert T_{\tau}
\left[ a_{\bf k}(\tau) a^{\dagger}_{\bf k}(0) 
\exp 
\left\{ 
- \int_{0}^{\infty} \! \hat{H}_{\mbox{\scriptsize int}}(\tau') d \tau'   
\right\}
\right]  
\right\vert \mbox{vac} \right\rangle_{\mbox{\scriptsize con}}
\; ; \; \tau > 0 \; .
\label{Gdefinition}
\end{equation}  
Here $T_{\tau}$ is the imaginary time ordering operator,  
$\vert \mbox{vac} \rangle$ is a vacuum state without particle and phonons,
$\hat{H}_{\mbox{\scriptsize int}}$ is the interaction Hamiltonian in (\ref{9}).
Symbol of exponent denotes Taylor expansion which results in multiple 
integration over internal variables  $ \{ \tau_1', \tau_2', \ldots \}$. 
Operators are in the interaction representation 
$\hat{A}(\tau) = 
\exp[\tau(\hat{H}_{\mbox{\scriptsize par}}+\hat{H}_{\mbox{\scriptsize ph}})] 
\hat{A}
\exp[-\tau(\hat{H}_{\mbox{\scriptsize par}}+\hat{H}_{\mbox{\scriptsize ph}})]
$.   
Index ``con'' means that expansion contains only connected terms
where no one integral over internal time variables 
$ \{ \tau_1', \tau_2', \ldots \}$ can be factorized. 

Vick theorem expresses matrix element of time-ordered operators as a sum 
of terms, each is a factor of matrix elements of pairs of operators, and
expansion  (\ref{Gdefinition}) becomes an infinite series of integrals 
with an ever increasing number of integration variables 
\begin{equation}
G_{\bf k} (\tau) \, = \, \sum_{m=0,2,4 \ldots}^{\infty} 
\sum_{\xi_m} \int dx_1' \cdots dx_m' \, 
{\cal D}_{m}^{(\xi_m)} (\tau ; \{ x_1', \ldots , x_m' \}) \; .
\label{main}
\end{equation}
Here index $\xi_m$ stands for different Feynman diagrams (FDs) 
of the same order $m$. 
Term with $m=0$ is the GF of the noninteracting QP
$G_{\bf k}^{(0)}(\tau)$. 
Function ${\cal D}_{m}^{(\xi_m)} (\tau ; \{ x_1', \ldots , x_m' \})$ 
of any order $m$ can be expressed as a factor of  
GFs of noninteracting quasiparticle, GFs of phonons, and interaction 
vortexes $V({{\bf k},\bf q})$. 
For the simplest case of Hamiltonian system expressions for GFs of 
QP 
$G_{{\bf k}}^{(0)}(\tau_2-\tau_1) = 
\exp \left[ -\epsilon({\bf k}) (\tau_2-\tau_1) \right]$ ($\tau_2 > \tau_1$)
and phonons
$D_{\bf q}^{(0)}( \tau_2-\tau_1) =
\exp \left[ -\omega_{\bf q} (\tau_2-\tau_1) \right]$ ($\tau_2 > \tau_1$)
are well known. 

\begin{figure}[h]
\hspace{-0.45 cm}  \vspace {-0.5 cm}
\includegraphics{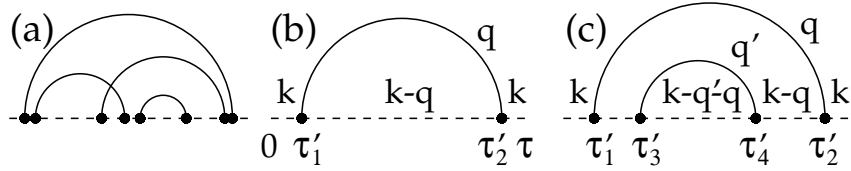}
\caption{\label{fig:DMC_1} (a) Typical FD contributing into
expansion (\ref{main}). (b) FD of the second order and (c) forth order.}
\end{figure}

An important feature of the DMC method, which is distinct from the row of 
other exact numerical approaches, is the explicit possibility to include 
renormalized GFs into exact expansion without any change of the algorithm. 
For example, if a damping of QP, caused by some interactions not included in the
Hamiltonian, is known, i.e. retarded self-energy of QP 
$\Sigma_{\mbox{\scriptsize ret}}({\bf k},\omega)$ is available, 
renormalized GF 
\begin{equation}  
\widetilde{G}_{{\bf k}}^{(0)}(\tau) =
\frac{1}{\pi}
\int_{-\infty}^{\infty} d \omega e^{- \omega \tau}
\frac{Im \Sigma _{\mbox{\scriptsize ret}}({\bf k},\omega) }
{\left[ \omega - \epsilon({\bf k}) - 
Re \Sigma _{\mbox{\scriptsize ret}}({\bf k},\omega) \right]^2 
+ \left[ Im \Sigma _{\mbox{\scriptsize ret}}({\bf k},\omega \right]^2}
\label{renora}
\end{equation}
can be introduced instead of bare GF $G_{{\bf k}}^{(0)}(\tau)$. 
Explicit rules for evaluation of ${\cal D}_{m}^{(\xi_m)}$ do not 
depend on the order and topology of FD. GFs of noninteracting
QPs $G_{{\bf k}}^{(0)}(\tau_2-\tau_1)$ 
(or $\widetilde{G}_{{\bf k}}^{(0)}(\tau_2-\tau_1)$) with corresponding
times and momenta are ascribed to horizontal lines and noninteracting
GFs of phonon $D_{\bf q}^{(0)}( \tau_2-\tau_1)$ 
(multiplied by the factor of corresponding vortexes
$V({\bf k}', {\bf q}) V^{*}({\bf k}'', {\bf q})$) 
are attributed to phonon propagator arch (see Fig.~\ref{fig:DMC_1}a).
Then, ${\cal D}_{m}^{(\xi_m)}$ is the factor of all GSs. 
For example, expression for the weight of the second order term
(Fig.~\ref{fig:DMC_1}b) is the following
\begin{eqnarray}
&{\cal D}_2 (\tau ; \{ \tau_2', \tau_1', {\bf q} \}) =& \nonumber \\
&\left\vert V({\bf k},{\bf q}) 
\right\vert^2 D^{(0)}_{{\bf q}} (\tau_2'-\tau_1') 
G^{(0)}_{{\bf k}} (\tau_1') G^{(0)}_{{\bf k-q}} (\tau_2'-\tau_1')
G^{(0)}_{{\bf k}} (\tau-\tau_2') \; .& 
\label{w-2}
\end{eqnarray}
%%%%%%%%%%%%%%%%%%%%%%%%%%%%%%%%%%%%%%%%%
%%%%%%%%%%%%%%%%%%%%%%%%%%%%%%%%%%%%%%%%%

The DMC process is a numerical procedure which, basing on the 
Metropolis principle \cite{Metro,LanBin}, samples different 
FDs in the parameter space $(\tau,m,\xi_m,\{ x_m' \} )$ and collects
statistics of external variable $\tau$ in a way that the result 
of this statistics converges to exact GF $G_{\bf k} (\tau)$. Although
sampling of the internal parameters of one term in (\ref{main}) 
and switch between different orders is performed within the 
the framework of one and the same numerical process, it is instructive 
to start with the procedure of evaluation of a specific term 
${\cal D}_{m}^{(\xi_m)} (\tau ; \{ x_1', \ldots , x_m' \})$.

Starting from a set 
$\{ \tau ; \{ x_1', \ldots , x_m' \} \}$, an update 
$x_l^{(old)} \rightarrow x_l^{(new)}$ of an arbitrary chosen 
parameter is suggested.  This update is accepted or rejected according 
to Metropolis principle. After many steps, altering all variables, 
statistics of external variable converges to exact dependence 
of the term on $\tau$.  
Suggestion for new value of parameter $x_l^{(new)} = \hat{S}^{-1}(R)$ is 
generated by random number $R \in [0,1]$ with a normalized  
distribution function $W(x_l)$ in a range 
$x_l^{(min)} < x_l < x_l^{(max)}$. There are only two restrictions for this
otherwise arbitrary function. First, new parameters $x_l^{(new)}$
must not violate FD topology, i.e., for example, internal time 
$\tau_1'$ in Fig.~\ref{fig:DMC_1}c must be in the range
$[x^{(min)}=0,x^{(max)}=\tau_3']$.
Second, the distribution must be nonzero for the whole, allowed by FD topology,
domain. This ergodicity property is crucial since it is necessary to 
sample the whole domain for convergence to exact answer.
At each step, update $x_l^{(old)} \rightarrow x_l^{(new)}$ is accepted 
with probability $P_{acc}=M$ (if $M<1$) and always otherwise.
The ratio $M$ has the following form
\begin{equation}  
M = 
\frac
{
{\cal D}_{m}^{(\xi_m)} 
(\tau ; \{ x_1', \ldots , x_l^{(new)}, \ldots , x_m' \})
/
W(x_l^{(new)})
}
{
{\cal D}_{m}^{(\xi_m)} 
(\tau ; \{ x_1', \ldots , x_l^{(old)}, \ldots , x_m' \})
/
W(x_l^{(old)})
} \; .
\label{ratio}
\end{equation}
For uniform distribution   
$W=\mbox{const}=\left( x_l^{(max)}-x_l^{(max)} \right)^{-1}$,
the probability of any combination of parameters is 
proportional to the weight function ${\cal D}$. 
However, for better convergence the distribution $W(x_l^{new})$ 
must be as close as possible to the actual distribution given by 
function ${\cal D}_{m}^{(\xi_m)} 
(\{ \ldots , x_l^{(new)}, \ldots , \})$.

For sampling over FDs of all orders and topologies it is enough 
to introduce two complimentary updates. Update ${\cal A}$ transforms 
FD  ${\cal D}_{m}^{(\xi_m)} (\tau ; \{ x_1', \ldots , x_m' \})$
into higher order FD
 ${\cal D}_{m+2}^{(\xi_{m+2})} 
(\tau ; \{ x_1', \ldots , x_m' ; \; {\bf q}', \tau'_3, \tau'_4 \})$
with extra phonon arch, connecting some time points 
$\tau'_3$ and $\tau'_4$ by phonon propagator with momentum ${\bf q}'$
(Fig.~\ref{fig:DMC_1}c).
Note that the ratio of weights 
${\cal D}_{m+2}^{(\xi_{m+2})} / {\cal D}_{m}^{(\xi_m)}$
is not dimensionless. Dimensionless Metropolis ratio   
\begin{equation}
M = \frac{p_{\cal A}}{p_{\cal B}}
\frac
{
{\cal D}_{m+2}^{(\xi_{m+2})} 
(\tau ; \{ x_1', \ldots , x_m' ; \; {\bf q}', \tau', \tau'' \})
}
{
{\cal D}_{m}^{(\xi_m)} (\tau ; \{ x_1', \ldots , x_m' \})
W( {\bf q}', \tau', \tau'' )
} \; .
\label{ratio2}
\end{equation}
contains normalized probability function
$W( {\bf q}', \tau', \tau'' )$, which is used 
for generating of new parameters\footnote{The factor $p_{\cal A}/p_{\cal B}$ 
depends on the probability to address add/remove processes.}. 
Complementary update ${\cal B}$, removing 
the phonon propagator, uses ratio $M^{-1}$ \cite{MPSS}.

Note that all updates are local, i.e. do not 
depend on the structure of the whole FD. 
Neither rules nor CPU time, needed for update, depends on the FD order. 
DMC method does not imply any explicit truncation of FDs order due to 
finite size of computer memory. 
Ever for strong coupling, where typical number of phonon
propagators $N_{ph}$, contributing to result, is large, influence
of finite size of memory is not essential. Really, according 
to Central Limit Theorem, number of phonon propagators obeys 
Gauss distribution centered at $\bar{N}_{ph}$ with half width of the 
order of $\sqrt{\bar{N}_{ph}}$ \cite{Sandvik}.
Hence, if a memory for at least $2\bar{N}_{ph}$ propagators is reserved, 
diagram order hardly surpasses this limit.

\subsection{Stochastic Optimization Method}
\label{sec_2_subsec_2}

The problem of inverting of integral equation (\ref{Fr}) 
is an ill posed problem.
Due to incomplete noisy information about GF 
$G_{\bf k}(\tau)$, which is known with statistic errors on a finite number of 
imaginary times in a finite range $[0,\tau_{\mbox{\scriptsize max}}]$, 
there is infinite number of approximate solutions
which reproduce GF within some range of deviations and the problem is 
to chose ``the best one''. Another problem, 
which is a stumbling block for decades, is the saw tooth noise instability.
It occurs when solution is obtained by a naive method, e.g. by using
least-squares approach for minimizing deviation measure 
\begin{equation}
D[\tilde{L}_{\bf k}(\omega)] = 
\int_{0}^{\tau_{\mbox{\scriptsize max}}} 
\left| G_{\bf k}(\tau) -  \tilde{G}_{\bf k}(\tau) \right| G^{-1}_{\bf k}(\tau)
d \tau  \; .
\label{ap_sp5}
\end{equation}
Here $\tilde{G}_{\bf k}(\tau)$ is obtained from approximate LF
$\tilde{L}_{\bf k}(\omega)$ by applying of integral operator
$\tilde{G}_{\bf k}(\tau) = {\cal F} \left[ \tilde{L}_{\bf k}(\omega) \right]$
in (\ref{Fr}).
Saw tooth instability corrupts LF in the ranges where actual LF is smooth. 
Fast fluctuations of the solution $\tilde{L}_{\bf k}(\omega)$ often have 
much larger amplitude than the value of actual LF $L_{\bf k}(\omega)$. 
Standard tools for saw tooth noise suppression are based on
the early 60-es idea of Fillips-Tikhonov regularization method 
\cite{regularization,regcm,Phillips,Tikhonov}.
A nonlinear functional, which suppresses large derivatives of approximate 
solution $\tilde{L}_{\bf k}(\omega)$, is added to the linear deviation 
measure (\ref{ap_sp5}).
Most popular variant of regularization methods is the Maximal Entropy 
Method \cite{JG}. 

However, typical LF of a QP in a boson field consists of 
$\delta$-functional peaks and smooth incoherent continuum with a
sharp border \cite{MPSS,RP02}.
Hence, suppression of high derivatives, as a general strategy of the 
regularization method, fails. Moreover, any specific implementation of 
the regularization method uses predefined mesh in the 
$\omega$ space, which could be absolutely unacceptable for the case 
of sharp peaks. If the actual location of a sharp peak is between 
predefined discrete points, the rest of spectral density can be 
distorted beyond recognition. Finally, regularization 
Maximal Entropy approach requires assumption of Gauss distribution 
of statistic errors in  $G_{\bf k}(\tau)$, which might be invalid 
in some cases \cite{JG}.

Recently, a Stochastic Optimization (SO) method,  
which circumvents abovementioned difficulties, was developed \cite{MPSS}. 
The idea of the SO method is to generate a large enough number
$M$ of statistically independent nonregularized solutions 
$\{ \tilde{L}_{\bf k}^{(s)}(\omega)\}, s=1,...,M$, which deviation 
measures $D^{(s)}$ are smaller than some upper limit $D_u$, depending of 
the statistic noise of the GF $G_{\bf k}(\tau)$.
Then, using linearity of expressions (\ref{Fr}), (\ref{ap_sp5}), 
the final solution is found as the average of particular solutions 
$\{ \tilde{L}_{\bf k}^{(s)}(\omega)\}$
\begin{equation}
L_{\bf k}({\omega}) \, = \, 
M^{-1} \sum_{s=1}^{M} \tilde{L}_{\bf k}^{(s)}(\omega)  \; .
\label{ap_sp4}
\end{equation}
Particular solution $\tilde{L}_{\bf k}^{(s)}(\omega)$ 
is parameterized in terms of sum
\begin{equation}
\tilde{L}_{\bf k}^{(s)}(\omega) = \sum_{t=1}^{K} \chi_{\{ P_t \}}(\omega)
\label{ap_sp6}
\end{equation}
of rectangles $\{ P_t \} =  \{ h_t,w_t,c_t \}$ with height $h_t>0$, 
width $w_t>0$, and center $c_t$. 
Configuration
\begin{equation}
{\cal C} = \left\{  {\{ P_t \}}, \, t=1, ... , K  \right\} \; ,
\label{ap_sp8}
\end{equation}
which satisfies normalization condition 
$\sum_{t=1}^{K} h_t w_t = 1$, 
defines function $\tilde{G}_{{\bf k}}(\tau)$. 
The procedure of generating particular solution starts from 
stochastic choice of initial configuration 
${\cal C}_s^{\mbox{\scriptsize init}}$. Then, deviation 
measure is optimized by a randomly chosen consequence of updates 
until deviation is less than $D_u$. In addition to updates, which 
do not change number of terms in the sum (\ref{ap_sp6}), there are 
updates which increase or decrease number $K$. Hence, since the number 
of elements $K$ is not fixed, any spectral function can be reproduced 
with desired accuracy.

Although each particular solution $\tilde{L}_{\bf k}^{(s)}(\omega)$
suffers from saw tooth noise at the area of smooth LF, 
statistical independence of each solution leads to 
self averaging of this noise in the sum (\ref{ap_sp6}). Note that 
suppression of noise happens without suppression of high derivatives and,
hence, sharp peaks and edges are not smeared out in contrast to 
regularization approaches. Therefore, saw tooth noise instability is 
defeated without corruption of sharp peaks and edges. Moreover, 
continuous parameterization (\ref{ap_sp6}) does not need predefined 
mesh in $\omega$-space. Besides, since the Hilbert space 
of solution is sampled directly, any assumption about distribution of
statistical errors is not necessary.  
 
SO method was successfully applied to restore LF of Fr\"{o}hlich 
polaron \cite{MPSS}, Rashba-Pekar exciton-polaron \cite{RP02},
hole-polaron in $t$-$J$-model \cite{Hole01,tJpho}, and many-particle
spin system \cite{Aplesnin03}.
Calculation of the optical conductivity 
of polaron by SO method can be found in \cite{Optics}. 
SO method appeared to be helpful in cases when GF's asymptotic limit, 
giving information about ground state, can not be reached. 
For example, sign fluctuations of the terms in expansion (\ref{main})
for a hole in the $t$-$J$-model lead to poor statistics at large times
\cite{Hole01}, though, SO method is capable of recovering energy 
and $Z$-factor even from GF known only at small imaginary times 
\cite{Hole01}.

\subsection{Advantages and Drawbacks of DMC-SO Machinery}
\label{sec_2_subsec_3}
   
Among numerical methods, capable of obtaining quantitative results in the 
problem of exciton (\ref{1}) and (\ref{2}), one can list 
time-dependent density functional theory \cite{Onida02}, 
Hanke-Sham technique of correcting particle-hole excitation energy
\cite{HanSham65,Benedict98},  and approaches directly solving 
Bethe-Salpeter equation \cite{Alb98,RohLo98,Marini03}. 
The latter ones provide rather accurate information on the two-particle GF. 
However, usage of finite mesh in direct/reciprocal space, 
which is avoided in DMC method, leads to its' failure in  
Wannier regime \cite{RohLo98}.    

In contrast to DMC method, none of the traditional numeric
methods can give reliable results for measurable properties of 
excited states of polaron at arbitrary range of electron-phonon interaction 
for the macroscopic system in the thermodynamic limit. 
Exact diagonalization method \cite{Ste96,WelFe97,FeLoWe97,FeLoWe97a} can 
study excited states though only on rather small finite size systems and 
results of this method are not even justified in the variational sense in 
the thermodynamic limit \cite{BoTruBa99}.
There is a batch of rather effective variational ``exact translation''
methods \cite{BoTruBa99,ChungTrug,ShaTrug,Barisic02,Barisic04} where 
basis is chosen in the momentum space and, hence, 
the variational principle is applied in the thermodynamic limit. 
Although these methods can reveal few discrete excited states, its fail
for long-range interaction and for dispersive, especially acoustic 
phonons due to catastrophic growth of variational basis. 
A non perturbative theory, which is able to give information about spectral
properties in the thermodynamic limit  at least for one electron, is 
Dynamical Mean Field Theory \cite{DMF1,DMF2,DMF3,DMF4}. 
However it gives an exact solution only in the case of
infinite dimension which does not correspond to a realistic system and can 
be considered only  as a guide for extrapolation to finite dimensions 
\cite{Chuki97}.

Recently developed cluster perturbation theory, where exact diagonalization 
of a cluster is further improved by taking into account inter-cluster
interaction \cite{Sene,Sene1,Hohen,Hohen1,Hohen2}, 
is applicable for study of the excited states,
but limited to one-dimensional lattices or two-dimensional systems 
with short-range interaction. Traditional density-matrix renormalization
group method \cite{Jeckel,Jeckel1,Jeckel2,Jeckel3,Jeckel4} 
is very effective though mostly limited to 
one-dimensional systems and ladders. Finally, recently developed 
path integral quantum Monte Carlo algorithm 
\cite{Kornilo,Kornilo1,Kornilo2,HagueKonilo06} is valid for any dimension 
and properly takes into account quasi long-range interactions 
\cite{AlexKornilo99}. Path integral method is capable of obtaining
the density of states \cite{Kornilo,Kornilo1} and isotope exponents 
\cite{Kornilo2,KorniloAlex04}. 
However calculations of measurable characteristics of excited states, 
such as ARPES or optical conductivity, by this method were never reported.   

In conclusion, none of methods, except DMC-SO combination, can obtain at 
the moment approximation-free results for {\it measurable physical quantities}
for a few QPs interacting with a macroscopic bosonic bath  
{\it in the thermodynamic limit}. 
Indeed, there are limitations of the DMC and SO methods.
DMC method does not work in many-fermion systems due 
to sign problem and SO method fails at high temperatures, comparable 
to energies of dominant spectral peaks, because even very small 
statistical noise of GFs turns Fredholm equation (\ref{Fr}) 
into essentially ``ill defined'' problem \cite{regularization}.

\section{Spectral Properties of the Fr\"{o}hlich Polaron}
\label{section_3}

Before development of DMC-SO methods, the information on the excited states of 
polaron models, especially the Fr\"{o}hlich one, was very limited. 
Knowledge of LF was based on results of infinite-dimensions approximation 
\cite{CiPaFe95}, exact diagonalization 
\cite{AleKaRay94,WelFe97,FeLoWe97,FeLoWe97}, or strong coupling expansion 
\cite{AlexRan}. 
No one of the above techniques was capable of obtaining the LF of polaron 
without approximations, especially for long-range interaction where 
difficulties of traditional numerical methods dramatically increase. 
In a similar way, optical conductivity (OC) of Fr\"{o}hlich model was known 
only in strong coupling expansion approximation \cite{LandPekar}, within the 
framework of the perturbation theory \cite{Gurevich}, or was based on the 
variational Feynman path integral technique \cite{Dev72}. In this sect.\ 
we consider exact DMC-SO results on LF \cite{MPSS} and OC 
\cite{Optics,FrCondon} of Fr\"{o}hlich polaron model.

\subsection{Lehmann Function of the Fr\"{o}hlich Polaron}
\label{sec_3_subsec_1}

The perturbation theory expression for the high-energy part 
($\omega>0$) of the LF for arbitrary interaction potential 
$V(\mid {\bf q} \mid)$ reads \cite{MPSS} (frequency of the optical phonon
$\omega_{ph}$ is set to unity) 
\begin{equation}
L_{{\bf k} =  0} (\omega > 0) = 
\frac{1}{\sqrt{2}\pi^2}
\frac{\sqrt{\omega-1}}{\omega^2} 
\mid V(\sqrt{2(\omega-1)}) \mid^2 \theta(\omega-1) \; .
\label{spa_2} 
\end{equation} 
%%%%%%%%%%%%%%%%%%%%%%%%%%%%%%%%%%%%%%%%%
\begin{figure}[h]
\hspace{-0.45 cm}  \vspace {-0.5 cm}
\includegraphics{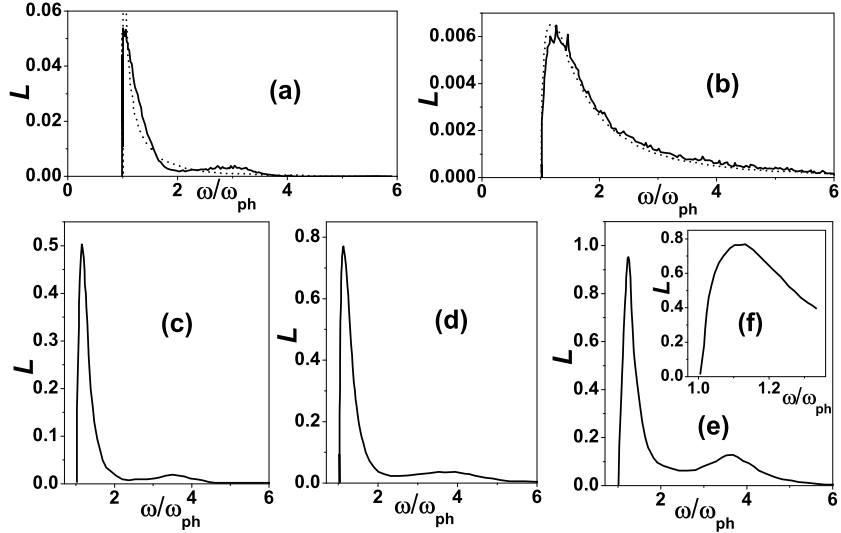}
\caption{\label{fig:al_fr_weak}
Comparison of the numerical results (solid lines) and the 
perturbation theory (dashed lines) for the LFs of 
the Fr\"{o}hlich model with $\alpha=0.05$ (a) and the short-range 
interaction model with $\alpha=0.05$ and $\kappa=1$ (b). 
LFs of Fr\"{o}hlich polaron for $\alpha=0.5$ (c), $\alpha=1$ (d) and 
$\alpha=2$ (e).
Energy is measured from that of the ground state of the polaron. 
The initial fragment of the LF for $\alpha=1$ is shown in the   
inset (f).}
\end{figure}
%%%%%%%%%%%%%%%%%%%%%%%%%%%%%%%%%%%%%%%%%
Low-energy part of the LF for the short-range interaction
%\begin{equation}
$V({\mid \bf q} \mid) \, = \, i \, 
\left( 2 \sqrt{2} \alpha \pi \right)^{1/2} \,  
(q^2+\kappa^2)^{-1/2}$ ,
%\end{equation}
reducing to the Fr\"{o}hlich one when $\kappa \to 0$, is
\begin{equation}
L_{{\bf k} = 0} (\omega < 0) = 
\frac{\alpha}{(\kappa + \sqrt{2})^2}
\delta \left( \omega + \alpha \frac{\sqrt{2}}{\kappa + \sqrt{2}} \right) \; . 
\label{spa_4} 
\end{equation}

Comparison of low-energy parts of the LF of the Fr\"{o}hlich model, 
obtained by DMC-SO and taken from (\ref{spa_4}), shows perfect agreement for 
$\alpha=0.05$:  the accuracy for the polaron energy and Z-factor is 
about $10^{-4}$. 
On the other hand, high-energy part of numeric result 
(Fig.~\ref{fig:al_fr_weak}) significantly deviates from that of the analytic 
expression (\ref{spa_4}). 
This is not surprising since for Fr\"{o}hlich polaron the perturbation 
theory expression is diverging as $\omega \to \omega_{ph}$ and, therefore 
the perturbation theory breaks down.
When perturbation theory is obviously valid, e.g.\ for the case of finite 
$\kappa=1$, there is a perfect agreement between analytic expression and 
DMC-SO results (Fig.~\ref{fig:al_fr_weak}b). 
Note that the high-energy part of $L_{{\bf k} = 0}(\omega )$
is successfully restored by SO method despite the fact that the total weight 
of the feature for $\alpha =0.05$ is less than $10^{-2}$.
%%%%%%%%%%%%%%%%%%%%%%%%%%%%%%%%%%%%%%%%%
\begin{figure}[h]
\hspace{-0.45 cm}  \vspace {-0.5 cm}
\includegraphics{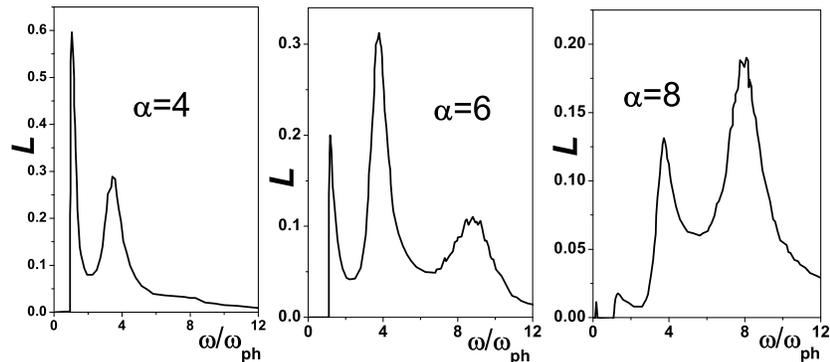}
\caption{\label{fig:al_fr_str} Evolution of spectral density with $\alpha $
in the cross-over region from intermediate to strong couplings. The 
polaron ground state peak is shown only for $\alpha=8$.
Note that the spectral analysis still resolves it, despite its very 
small weight $<10^{-3}$. }
\end{figure}
%%%%%%%%%%%%%%%%%%%%%%%%%%%%%%%%%%%%%%%%%

The main deviation of the actual LF from the perturbation theory result is the
extra broad peak in the actual LF at $\omega \sim 3.5$. To study this 
feature $L_{{\bf k} = 0}(\omega)$ was calculated for $\alpha=0.5$, $\alpha=1$,
and $\alpha=2$ (Fig.~\ref{fig:al_fr_weak}c-e). 
The peak is seen for higher values of the interaction constant and its weight 
grows with $\alpha$. 
Near the threshold, $\omega=1$, LF demonstrates the square-root dependence 
$\sim \sqrt{\omega-1}$ (Fig.~\ref{fig:al_fr_weak}f).

To trace the evolution of the peak at higher values of $\alpha$ the LF was 
calculated \cite{MPSS} for $\alpha=4$, $\alpha=6$, and $\alpha=8$ 
(Fig.~\ref{fig:al_fr_str}). 
At $\alpha=4$ the peak at $\omega \sim 4$ already dominates.
Moreover, a distinct high-energy shoulder appears at $\alpha=4$,
which transforms into a broad peak at $\omega \sim 8.5$ in the
LF for $\alpha=6$. 
The LF for $\alpha=8$ demonstrates further redistribution of the spectral 
weight between different maxima without significant shift of the peak 
positions.

\subsection{Optical Conductivity of the Fr\"{o}hlich Polaron: Validity
of the Franck-Condon Principle in the Optical Spectroscopy}
\label{sec_3_subsec_2}

The FC principle \cite{Franck,Condon} and its validity have been widely 
discussed in studies of optical transitions in atoms, molecules 
\cite{Bertran,Urbain}, and solids \cite{Lax,Toyobook}.
Generally, the FC principle means that if only one of two coupled subsystems, 
e.g.\ an electronic subsystem, is affected by an external perturbation, the
second subsystem, e.g., the lattice, is not fast enough to follow the 
reconstruction of the electronic configuration. 
It is clear that the justification for the FC principle is the short 
characteristic time of the measurement process $\tau_{mp} \ll \tau_{ic}$, 
where $\tau_{mp}$ is related to the energy gap between the initial and final 
states, $\Delta E$, through the uncertainty principle: 
$\tau_{mp}\simeq \hbar / (\Delta E)$ and $\tau_{ic}$ is the time 
necessary to adjust the lattice when the electronic component is perturbed. 
Then, the spectroscopic response considerably depends on the value of the 
ratio $\tau_{mp}/\tau_{ic}$  
For example, in mixed valence systems, where the ionic valence fluctuates 
between configurations $f^5$ and $f^6$ with characteristic time  
$\tau_{ic} \approx 10^{-13}$s, spectra of fast and slow experiments are 
dramatically different \cite{Khomsky,Falicov}. 
Photoemission experiments with short characteristic times 
$\tau_{mp} \approx 10^{-16}$s (FC regime), 
reveal two lines, corresponding to $f^5$ and $f^6$ states.  
On the other hand slow M\"ossbauer isomer shift measurements with 
$\tau_{mp} \approx 10^{-9}$s show a single broad peak 
with mean frequency lying between signals from pure $f^5$ and $f^6$ shells.
Finally, according to paradigm of measurement process time, 
magnetic neutron scattering with $\tau_{mp} \approx \tau_{ic}$ revealed 
both coherent lines with all subsystems dynamically adjusted and broad 
incoherent remnants of strongly damped excitation of $f^5$ and $f^6$ 
shells \cite{Aleks95,KiMi95}. 
Actually, the  meaning of the times $\tau_{ic}$  and $\tau_{mp}$ varies 
with the system and with the measurement process.

To study the interplay between measurement process time $\tau_{mp}$ and
adjustment time $\tau_{ic}$, the OC of the Fr\"ohlich polaron was studied in 
\cite{FrCondon} from the weak to the strong coupling regime by three 
methods. 
DMC method gives numerically exact answer which is compared with memory 
function formalism (MFF), which is able to take dynamical lattice relaxation 
into account, and strong coupling expansion (SCE) which assumes FC approach.
It was found that near critical coupling $\alpha_c \approx 8.5$ a dramatic 
change of the OC spectrum occurs: dominating peak of OC splits into two 
satellites.  In this critical regime the upper (lower) one quickly 
decreases (increases) it's spectral weight as the value of 
coupling constant increases.
Besides, while OC follows prediction of MFF at $\alpha<\alpha_c$, its 
dependence switches to that predicted by SCE for larger couplings.
It was concluded that, for the OC measurement of polaron, the adjustment time 
$\tau_{ic} \approx \hbar / {\cal D}$  is set by typical nonadiabatic energy 
${\cal D}$. 
Nonadiabaticity destroys FC classification at $\alpha<\alpha_c$ while FC 
principle rapidly regains its validity at large couplings due to fast growth 
of energy separation between initial and final states of optical transitions.  

Comparison of exact DMC-SO data for OC with existing results of approximate
methods showed \cite{Optics} that the Feynman path integral technique 
\cite{Dev72} of Devreese, De Sitter, and Goovaerts, where OC is calculated 
starting from the Feynman variational model \cite{Fey55}, is the only 
successfully describing evolution of the energy of the main peak in OC 
with coupling constant $\alpha$ (see \cite{D}). 
However, starting from the intermediate coupling regime this approach fails 
to reproduce the peak width. Subsequently, the path integral approach was 
rewritten in terms of MFF \cite{PeetersDev83}. 
Then, in \cite{FrCondon} the extended MFF formalism, which 
introduces dissipation processes fixed by exact sum rules, was developed
\cite{Cataudella}.    
  
\begin{figure}[h]
\hspace{-0.45 cm}  \vspace {-0.5 cm}
\includegraphics{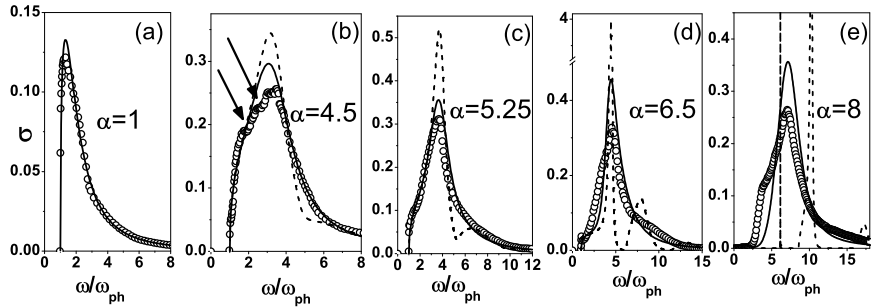}
\caption{\label{fig:al_fro_opt} Comparison of the optical conductivity 
calculated by DMC method (circles), extended MFF (solid line), and DSG 
\cite{Dev72,PeetersDev83} (dashed line) for different values of $\alpha$.
The slanted arrows indicate 2- and 3-phonon thresholds of absorption.}
\end{figure}

As shown in Fig.~\ref{fig:al_fro_opt}a, in the weak coupling regime, the 
MFF, with or without dissipation, is in very good agreement with DMC data, 
showing significant improvement with respect to weak coupling perturbation 
approach \cite{Gurevich} which provides a good description of OC spectra 
only for very small values of $\alpha$. 
For $1 \le  \alpha \le 8$, where standard MFF fails to reproduce peak width
(Fig.~\ref{fig:al_fro_opt}b-d) and even the peak position 
(Fig.~\ref{fig:al_fro_opt}c), the damping, introduced to extended MFF scheme, 
becomes crucial. 
Results of extended MFF are accurate for the peak energy and quite 
satisfactory for the peak width (Fig.~\ref{fig:al_fro_opt}b-e).
Note that the broadening of the peak in DMC data is not a consequence of 
poor quality of analytic continuation procedure since DMC-SO methods 
is capable of revealing such fine features as 2- and 3-phonon thresholds
of emission (Fig.~\ref{fig:al_fro_opt}b).     

%%%%%%%%%%%%%%%%%%%%%%%%%%%%%%%%%%%%%%%%%
%%%%%%%%%%%%%%%%%%%%%%%%%%%%%%%%%%%%%%%%%

%%%%%%%%%%%%%%%%%%%%%%%%%%%%%%%%%%%%%%%%%
\begin{figure}[h]
\hspace{-0.45 cm}  \vspace {-0.5 cm}
\includegraphics{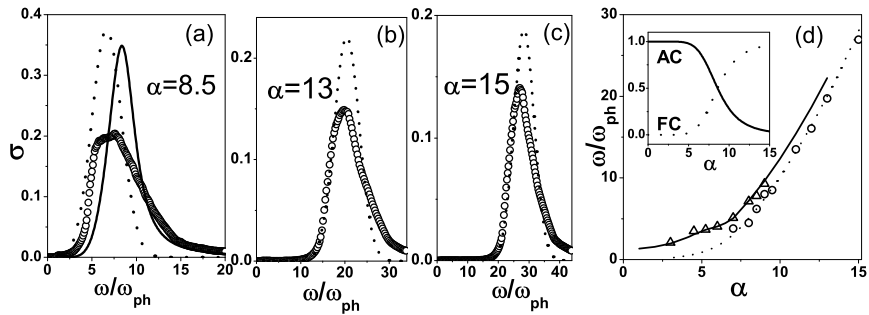}
\caption{\label{fig:al_fro_opt2} (a)-(c) Comparison of the optical 
conductivity calculated within the DMC method (circles), the extended MFF
(solid line), and SCE (dotted line) for different values of $\alpha$.
(d) The energy of lower- and higher-frequency features (circles and triangles, 
respectively) compared with the FC transition energy with the SCE 
(dashed line) and with the energy of the peak obtained from the extended MFF
(solid line). In the inset, the weights of FC and adiabatically 
connected transitions are shown as a function of $\alpha$ (for $\eta=1.3$.)}
\end{figure}
%%%%%%%%%%%%%%%%%%%%%%%%%%%%%%%%%%%%%%%%%

However, a dramatic change of OC occurs around critical coupling 
strength $\alpha_c \approx 8.5$. The dominating peak of OC splits 
into two ones, the energy of lover one corresponding to the predictions of SCR
expansion and that of upper one obeying extended MFF
value (Fig.~\ref{fig:al_fro_opt2}a). 
The shoulder, corresponding to dynamical extended MFF contribution,
rapidly decreases it's intensity with increase of $\alpha$ and at 
large $\alpha$ (Fig.~\ref{fig:al_fro_opt2}b-c) the OC is in good agreement 
with strong coupling expansion, assuming FC scheme. 
Finally, comparing energies of the peaks, obtained by DMC, extended MFF and
FC strong coupling expansion (Fig.~\ref{fig:al_fro_opt2}d), we conclude that 
at critical coupling $\alpha_c \approx 8.5$ the spectral properties rapidly 
switch from dynamic, when lattice relaxes at transition, to FC regime, 
where nuclei are frozen in initial configuration. 

In order get an idea of the FC breakdown authors of \cite{FrCondon} 
consider the following arguments. 
The approximate adiabatic states are not exact eigenstates of the system. 
These states are mixed by nondiagonal matrix elements of the nonadiabatic 
operator ${\cal D}$ and exact eigenstates are linear combinations of the 
adiabatic wavefunctions. 
Being interested in the properties of transition from ground ($g$) to 
an excited ($ex$) state, whose energy correspond to that of the OC peak, 
it is necessary to consider mixing of only these states and express exact 
wavefunctions as a linear combinations \cite{Brovman,KiMi}
of ground and excited adiabatic states. 
The coefficients of superposition are determined from standard techniques
\cite{Brovman,KiMi} where nondiagonal matrix elements of the nonadiabatic 
operator \cite{Brovman} are expressed in terms of matrix elements of the 
kinetic energy operator $M$, the gap between excited and ground state 
$\Delta E = E_{ex}-E_{g}$ and the number $n_{\beta}$ of phonons in adiabatic 
state: 
\begin{equation} 
{\cal D^{ \pm} }=
M (\Delta E)^{-1} \sqrt{n_{\beta} + 1/2 \pm 1/2}
+
M^{2} (\Delta E)^{-2}.
\end{equation} 
The extent to which lattice can follow transition between
electronic states, depends on the degree of mixing between initial
and final exact eigenstates through the nonadiabatic interaction. 
If initial and ground states are strongly mixed, the 
adiabatic classification has no sense and, therefore, the FC 
processes have no place and lattice is adjusted to the change of 
electronic states during the transition. 
In the opposite limit adiabatic approximation is valid and FC processes 
dominate. 
The estimation for the weight of FC component $I_{FC}$ \cite{FrCondon} is 
%\begin{equation}
%I_{FC} = 1 - 
%4 \left\vert \xi_{g,\beta}^{G} \xi_{ex,\beta'}^{EX} \right\vert^2 \; ,
%\end{equation}
equal to unity in the case of zero mixing and zero in the case of maximal 
mixing. The weight of adiabatically connected (AC) transition 
$I_{AC}=1-I_{FC}$ is defined accordingly. 
Non-diagonal matrix element $M$ is proportional to the root square of 
$\alpha$ with a coefficient $\eta$ of the order of unity.
In the strong coupling regime, assuming that 
$\Delta E \approx  \gamma \alpha^2$ ($\gamma \approx 0.1$ from MC data), 
and $n_{\beta} \approx \Delta E$ ($n_{\beta} \gg 1$), one gets  
\begin{equation} 
I_{FC} = \left[
1 + 4 (\tau_{mp}/\tau_{ic})^{2} 
\right]^{-1} \; ,
\end{equation}
where $\tau_{mp}= 1/\Delta E$ and $\tau_{ic}= 1/D$.   
For $\eta$ of the order of unity one obtains qualitative description of a 
rather fast transition from AC- to FC-dominated transition, 
when $I_{FC}$ and $I_{AC}$ exchange half of their weights in the range of 
$\alpha$ from 7 to 9. 
The physical reason for such quick change is the faster growth of energy 
separation $\Delta E \sim \alpha^2$ compared to that of the  
matrix element $M  \sim \alpha^{1/2}$. Finally, for large couplings, 
initial and final states become adiabatically disconnected. 
The rapid AC-FC switch has nothing to do with the self-trapping phenomenon 
where crossing and hybridization of the ground and an excited states occurs. 
This phenomenon is a property of transition between different states and 
related to the choice whether lattice can or can not follow adiabatically the 
change of electronic state at the transition.

\section{Self-Trapping}
\label{section_4}

In this section we consider the self-trapping (ST) phenomenon which, due 
to essential importance of many-particle interaction of QP with bosonic 
bath of {\it macroscopic system}, was never addressed by exact method before.
We start with a basic definition of the ST phenomenon and introduce the
adopted criterion for it's existence. 
Then, generic features of ST are demonstrated on a simple model of 
Rashba-Pekar exciton-polaron in Sect.\ \ref{sec_4_subsec_1}.  
It is shown in Sect.\ \ref{sec_4_subsec_2} that the criterion is 
not a dogma since even in one dimensional system, where ST is forbidden by 
criterion of existence, one can observe all main features of ST due to 
peculiar nature of electronic states.

In general terms \cite{UKKTH86,R82}, ST is a dramatic transformation of a 
QP properties when system parameters are slightly changed. 
The physical reason of ST is a quantum resonance, which happens at some 
critical interaction constant $\gamma_c$, between ``trapped'' (T) 
state of QP with strong lattice deformation around it and ``free'' 
(F) state. 
Naturally, ST transition is not abrupt because of nonadiabatic 
interaction between T and F states and all properties of the QP are 
analytic in $\gamma$ \cite{Gerlach}. 
At small $\gamma<\gamma_c$, ground state is an F state which is weakly 
coupled to phonons while excited states are T states and have a 
large lattice deformation. At critical couplings $\gamma \approx \gamma_c$
a crossover and hybridization of these states occurs. Then, for 
$\gamma>\gamma_c$ the roles of the states exchange. The lowest state is 
a T state while the upper one is an F state.   

First, and up to now the only quantitative criterion for ST 
existence was given in terms of the ground state properties
in the adiabatic approximation. 
This criterion considers stability of the delocalized state in undistorted  
lattice $\Delta = 0$ with respect to the energy gain due to lattice distortion
$\Delta' \ne 0$. 
ST phenomenon occurs when completely delocalized state with $\Delta = 0$
is separated from distorted state with $\Delta' \ne 0$ by a barrier of 
adiabatic potential. 
One of these states is stable while another one is meta-stable. 
The criterion of barrier existence is defined in terms of the stability index
\begin{equation}   
s=d-2(1+l) \; ,
\label{sindex} 
\end{equation}
where $d$ is the system dimensionality. Index $l$ determines the range of the
force $\lim_{q \to 0} \psi(q) \sim q^{-l}$, where 
$\psi({\bf R})$ is the kernel of interaction
$U({\bf R}_n) = \psi({\bf R}_n-{\bf R}_{n'}) \nu({\bf R}_{n'})$
connecting potential $U({\bf R}_n)$ with generalized lattice distortion 
$\nu({\bf R}_{n'})$ \cite{UKKTH86}. 
The barrier exists for $s>0$ and does not exist for $s<0$. 
The discontinuous change of the polaron state, i.e.\ ST, occurs 
in the former case while does not happen in the latter case. 
When $s=0$, this scaling argument alone can not conclude the presence 
or absence of the ST and more detailed discussion for each model is needed. 
%%%%%%%%%%%%%%%%%%%%%%%%%%%%%%%%%%%%%%%%%
\begin{figure}[thb]
\hspace{0.0 cm}  \vspace {-0.5 cm}
\includegraphics{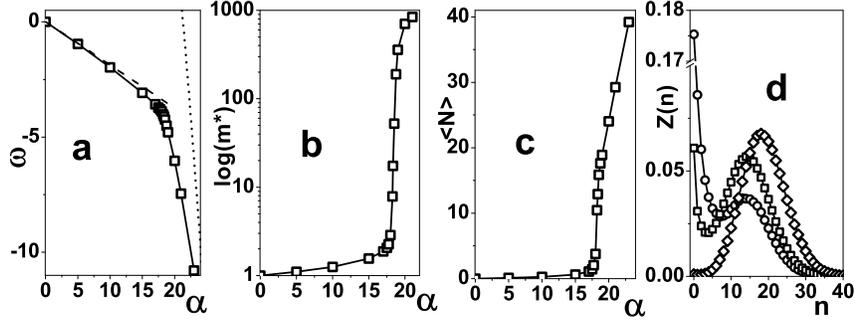}
\caption{\label{fig:var_rp_g} The ground-state energy (a), 
effective mass (b), and average number of phonons as function of 
coupling constant (c). 
Partial weights of $n$-phonon states (d) in the polaron ground state 
(${\bf k}=0$) at $\gamma=18$ (circles), $\gamma=18.35$ (squares), and
$\gamma=19$ (diamonds). Dotted line in panel (a) is the result of 
strong coupling limit and dashed line is the result of perturbation
theory.
} 
\end{figure}
%%%%%%%%%%%%%%%%%%%%%%%%%%%%%%%%%%%%%%%%%

\subsection{Typical Example of the Self-Trapping: Rasba-Pekar Exciton-Polaron}
\label{sec_4_subsec_1}

Classical example of a system with ST phenomenon is the three dimensional 
continuous Rasba-Pekar exciton-polaron in the approximation of 
intraband scattering, i.e. when polar electron-phonon interaction (EPI) 
with dispersionless optical phonons $\omega_{\mbox{\scriptsize ph}}=1$
does not change the wave function of internal electron-hole motion.
System is defined as a structureless QP with dispersion 
$\epsilon({\bf k})=k^2/2$ and short range coupling to phonons 
\cite{RP02,UKKTH86}. General criterion of the existence of ST is satisfied 
for three dimensional system with short range interaction 
\cite{RP02,UKKTH86,UFN_05} and, thus, one expects to observe typical 
features of the phenomenon.  

It is shown \cite{RP02} that in the vicinity of the critical coupling
$\gamma_c \approx 18$ the average number of phonons $\langle N \rangle$ in 
(\ref{srefon}) and effective mass $m^{*}$ quickly increase in the
ground state by several orders of magnitude (Fig.~\ref{fig:var_rp_g}b-c). 
Besides, a quantum resonance between polaronic phonon 
clouds of F and T state is demonstrated. 
Distribution of partial $n$-phonon contributions $Z^{({\bf k}=0)} (n)$ 
in (\ref{Zfactor}) has one maximum at $n=0$ in the weak coupling regime, which 
corresponds to weak deformation, and one maximum at $n \gg 1$ in the strong 
coupling regime, which is the consequence of a strong lattice distortion. 
However, due to F-T resonance there are two distinct peaks 
at $n=0$ and $n \gg 1$ for $\gamma \approx \gamma_c$
(Fig.~\ref{fig:var_rp_g}d).
%%%%%%%%%%%%%%%%%%%%%%%%%%%%%%%%%%%%%%%%%
\begin{figure}[htb]
\hspace{0.0 cm}  \vspace {-0.5 cm}
\includegraphics{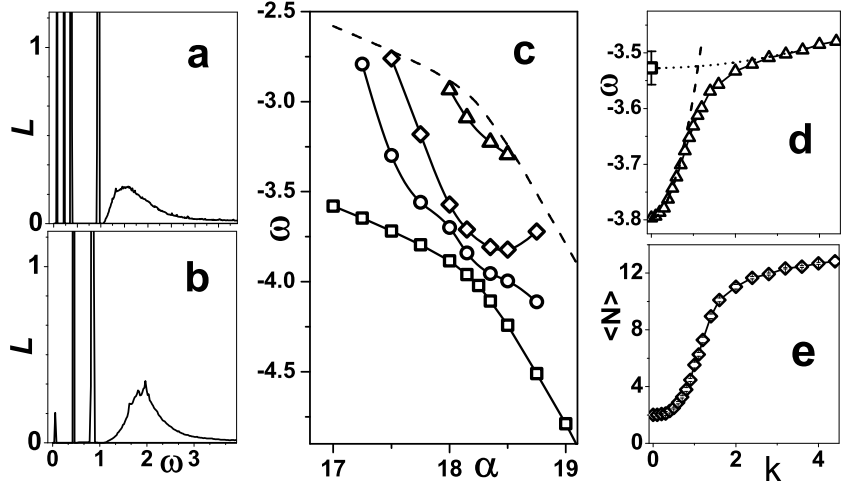}
\caption{\label{fig:var_rp} LF $L_{({\bf k}=0)}(\omega)$ at critical coupling
$\gamma=\gamma_c$ (a) and for $\gamma > \gamma_c$ (b). Energy is counted from 
the polaron ground state. (c) Dependence of energy of ground state (squares)
and stable excited states (circles, diamonds, and triangles) on the 
coupling constant. Dashed line is the threshold of the incoherent continuum.
Dependence of energy (d) and average number of phonons (e)
on the wave vector at $\gamma<\gamma_c$ (circles and rectangles). 
Dashed line is the 
effective mass approximation $E^{({\bf k})}= E_{gs} + {\bf k}^2/2m^*$ for
parameters $E_{gs}=-3.7946$ and $m^*=2.258$, obtained by DMC estimators for 
given value of $\gamma$. Dotted line is a parabolic dispersion law which is 
fitted to last 4 points of energy dispersion curve with parameters   
$E_{1}=-3.5273$ and $m^*_1=195$. Empty square is the energy of first 
excited stable state at zero momentum obtained by SO method.
} 
\end{figure}
%%%%%%%%%%%%%%%%%%%%%%%%%%%%%%%%%%%%%%%%%

Near the critical coupling $\gamma_c$ the LF of polaron 
has several stable states (Fig.~\ref{fig:var_rp} a-b) below 
the threshold of incoherent continuum 
$E_{\mbox{\scriptsize gs}}+\omega_{\mbox{\scriptsize ph}}$. Any state 
above the threshold is unstable because emission of a phonon with transition 
to the ground state at ${\bf k}=0$ with energy $E_{\mbox{\scriptsize gs}}$
is allowed.
On the other hand, decay is forbidden by conservation laws for states below
the threshold. 
Dependence of the energies of ground and excited resonances on the 
interaction constant resembles a picture of crossing of several
states interacting with each other (Fig.~\ref{fig:var_rp}c).    

According to the general picture of the ST phenomenon, lowest F state 
in the weak coupling regime at ${\bf k}=0$ has small effective mass 
$m^{*} \approx m$ of the order of the bare QP mass $m$. To the contrary, 
the effective mass of excited state $m^{*} \gg m$ is large. Hence, below 
the critical coupling the energy of the F state, which is lowest at
${\bf k}=0$, has to reach a flat band of T state at some momentum.
Then, F and T state have to hybridize and exchange in energy. 
DMC data visualize this picture (Fig.~\ref{fig:var_rp} d-e).
After F state crosses the flat band of excited T state, the average 
number of phonons increases and dispersion becomes flat.    

It is natural to assume that above the critical coupling the situation 
is opposite: ground state is the T state with large effective mass 
while excited F state has small, nearly bare, effective mass. 
Indeed, this assumption was confirmed in the framework of another 
model which is considered in Sect.\ \ref{sec_6_subsec_1}. 
Moreover, it was shown that in the strong coupling regime 
excited resonance inherits not only bare effective 
mass around ${\bf k}=0$ but the whole dispersion law of the bare 
QP \cite{tJpho}.

\subsection{Degeneracy Driven Self-Trapping}
\label{sec_4_subsec_2}

According to the criterion (\ref{sindex}), ST phenomenon in one-dimensional 
system does not occur. 
Although this statement is probably valid for the case of single band in 
relevant energy range, it is not the case for the generic multi-band cases. 
This fact has been unnoticed for many years, which prevented the proper 
explanation of puzzling physics of quasi-one-dimensional compound 
Anthracene-PMDA, although it's optical properties 
\cite{HaPhi75,BrPh80,Ha74,Ha77,ElsWeis85,KuGo94} 
directly suggested resonance of T and F states. 
The reason is that in Anthracene-PMDA, in contrast to conditions at which
criterion (\ref{sindex}) is obtained, there are two, nearly degenerate
exciton bands. Then, one can consider quasi-degenerate self-trapping
mechanism when ST phenomenon is driven by nondiagonal interaction 
of phonons with quasidegenerate exciton levels \cite{MN01}.
Such mechanism was already suggested for explanation of 
properties of mixed valence systems \cite{KiMi} though it's 
relevance was never proved by an exact approach.     

%%%%%%%%%%%%%%%%%%%%%%%%%%%%%%%%%%%%%%%%%
\begin{figure}[bht]
\begin{center}
\includegraphics[width=14cm]{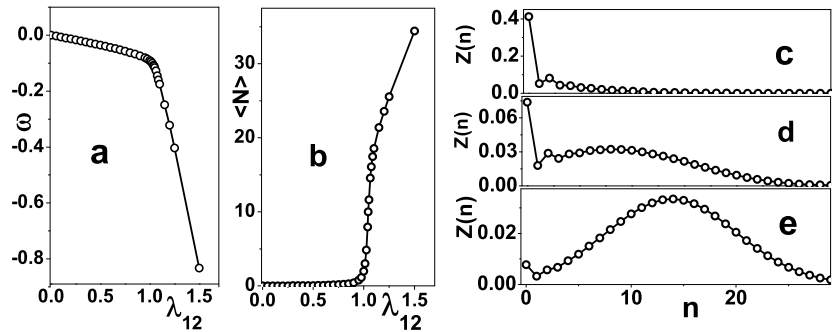}
\end{center}
\caption{ Dependence of energy (a) and average number of phonons (b)
on the nondiagonal coupling constant $\lambda_{12}$ at $\lambda_{11}=0$ 
and $\lambda_{22}=0.25$. Phonon distributions in polaron cloud 
below ST point at $\lambda_{12}=1.0125$ (c), at ST point at   
$\lambda_{12}=1.0435$ (d), and above ST coupling at 
$\lambda_{12}=1.0625$ (e).}
\label{var_odst}
\end{figure}
%%%%%%%%%%%%%%%%%%%%%%%%%%%%%%%%%%%%%%%%%
 
The minimal model to demonstrate the mechanism of quasi-degenerate
self-trapping involves one optical phonon branch with frequency 
$\omega_{ph}=0.1$ and two exciton branches with energies 
$\epsilon_{1,2}(q) = \Delta_{1,2} + 2[1-\cos(q)]$, 
where $\Delta_1=0$ and $\Delta_2=1$. 
Presence of short range diagonal $\gamma_{22}$ 
and nondiagonal $\gamma_{12}$ interactions 
(with corresponding dimensionless constants 
$\lambda_{22} = \gamma_{22}^2/(2\omega)$ and 
$\lambda_{12} = \gamma_{12}^2/(2\omega)$) leads to classical
self-trapping behavior even in one-dimensional system \cite{MN01} 
(see Fig.~\ref{var_odst}).

\section{Exciton}
\label{section_5}

Despite numerous efforts over the years, there has been no
rigorous technique to solve for exciton properties even for the
simplest model (\ref{1})-(\ref{2}) which treats electron-electron
interactions as a static renormalized Coulomb potential with
averaged dynamical screening. 
The only solvable cases are the Frenkel small-radius limit \cite{Frenkel} 
and the Wannier large-radius limit \cite{Wannier} which describe molecular 
crystals and wide gap insulators with large dielectric constant,
respectively. 
Meanwhile, even the accurate data for the limits of validity of the Wannier 
and Frenkel approximations have not been available. 
As discussed in Sects. \ref{sec_1_subsec_2} and \ref{sec_2_subsec_3}, 
semianalytic approaches has little to add to problem when quantitative 
results are needed whereas traditional numerical methods fail to reproduce 
them even in the Wannier regime. To the contrary, DMC results do not contain 
any approximation.

%%%%%%%%%%%%%%%%%%%%%%%%%%%%%%%%%%%%%%%%%
\begin{figure}[htb]
\hspace{0.0 cm}  \vspace {-0.5 cm}
\includegraphics[width=14cm]{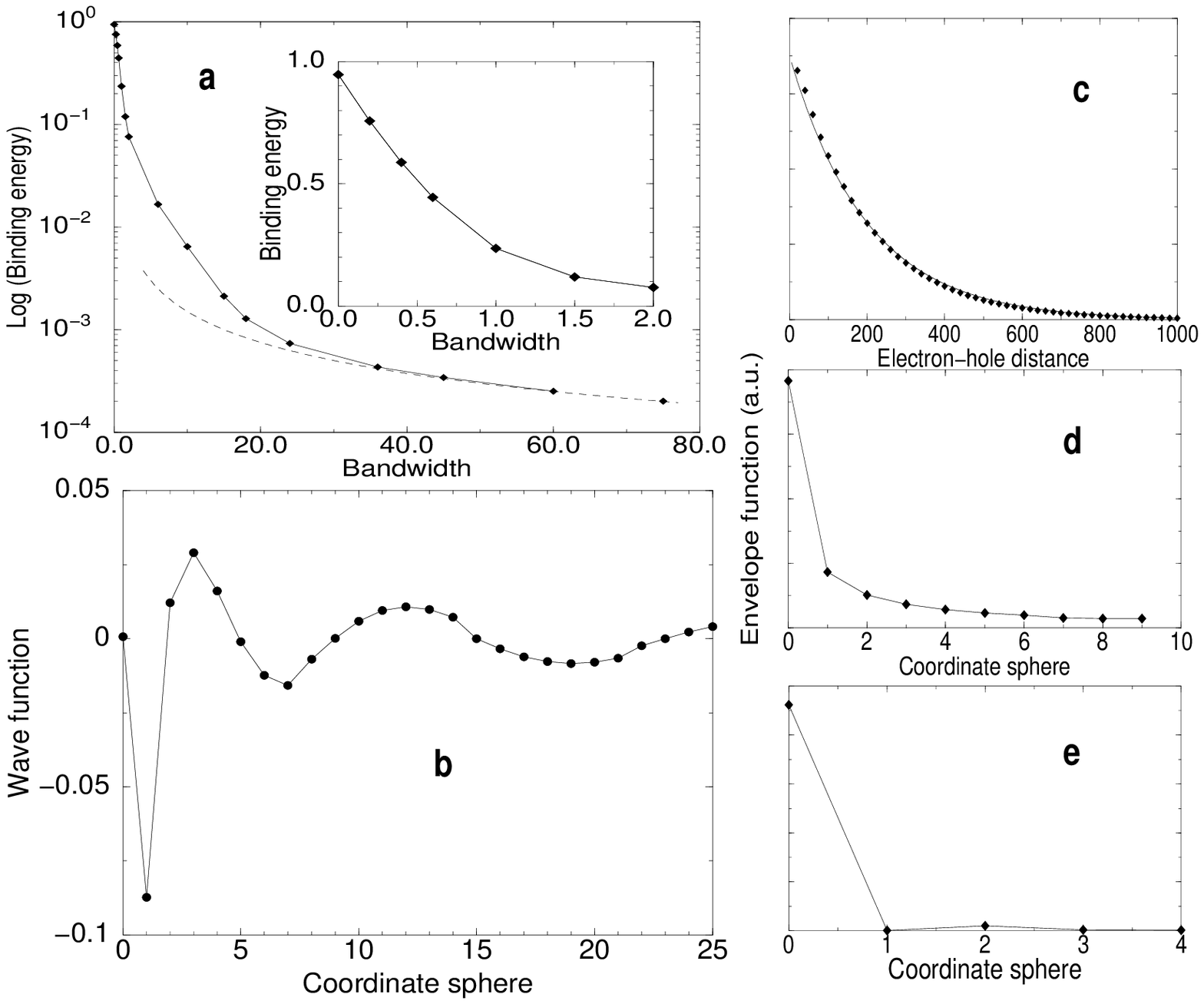}
\caption{\label{fig:var_exc} 
Panel (a): dependence of the exciton binding energy on the bandwidth $E_c=E_v$
for conduction and valence bands.
The dashed line corresponds to the Wannier model.  The
solid line is the cubic spline, the derivatives at the right and
left ends being fixed by the Wannier limit and perturbation
theory, respectively. Inset in panel (a): the initial part of the plot.
Panel (b): the wave function of internal motion in real space for the optically
forbidden monopolar exciton. 
Panels (c)-(e): the wave function of internal motion in real space: (c) Wannier
[$E_c=E_v=60$]; (d) intermediate [$E_c=E_v=10$]; (e) near-Frenkel
[$E_c=E_v=0.4 $] regimes.
The solid line in the panel (c) is the Wannier model result while solid 
lines in other panels are to guide the eyes only.}
\end{figure}
%%%%%%%%%%%%%%%%%%%%%%%%%%%%%%%%%%%%%%%%%

To study conditions of validity of limiting regimes by DMC method,
electron-hole spectrum of three dimensional system was chosen in the 
form of symmetric valence and conduction bands with width $E_c$ 
and direct gap $E_g$ at zero momentum \cite{Exciton}. 
For large ratio $W=E_c/E_g$, when $W>30$, exciton binding energy is 
in good agreement with Wannier approximation results 
(Fig.~\ref{fig:var_exc}a) and probability density of relative 
electron-hole motion corresponds (Fig.~\ref{fig:var_exc}c) to
hydrogen-like result. 
The striking result is the requirement of rather large valence and 
conduction bandwidths ($W>20$) for applicability of Wannier approximation. 
For smaller values of $W$ the binding energy and wave function
of relative motion (Fig.~\ref{fig:var_exc}d) deviate from large 
radius results. 
In the similar way, conditions of validity of Frenkel approach are rather 
restricted too. Moreover, even strong localization of wave function does 
not guarantee good agreement between exact and Frenkel approximation 
result for binding energy. 
At $1<W<10$ the wave function is already strongly localized though 
binding energy considerably differs from Frenkel approximation result.        
For example, at $W=0.4$ relative motion is well localized 
(Fig.~\ref{fig:var_exc}e) whereas the binding energy of Frenkel 
approximation is two times larger than exact result 
(Inset in Fig.~\ref{fig:var_exc}a). 

A study of conditions necessary for formation of charge transfer exciton in 
three dimensional systems is crucial to finalize protracted discussion 
of numerous models concerning properties of mixed valence semiconductors 
\cite{CuKiMi}. 
A decade ago unusual properties of SmS and SmB$_6$ were explained by invoking
the excitonic instability mechanism assuming charge-transfer
nature of the optically forbidden exciton \cite{KiMiZhe88,KiMi90}. 
Although this model explained quantitatively the phonon spectra 
\cite{Aleks89,KiMi91}, optical properties \cite{TraWa84,Le95}, and magnetic 
neutron scattering data \cite{KiMi95}, it's basic assumption has been 
criticized as being groundless \cite{Kas94_1,Kas94_2}. 
To study excitonic wavefunction, dispersions of valence and conduction bands 
were chosen as it is typical for mixed valence materials: almost flat 
valence band is separated from broad conduction band, having maximum in the 
centre and minimum at the border of Brillouin zone \cite{Exciton}. 
Results presented in Fig.~\ref{fig:var_exc}b support assumption of 
\cite{KiMiZhe88,KiMi90} since wave function of relative motion 
has almost zero on-site component and  maximal charge density at near 
neighbors.

\section{Polarons in Undoped High Temperature Superconductors}
\label{section_6}

It is now well established that the physics of high temperature 
superconductors is that of hole doping a Mott insulator 
\cite{ManoRev94,Dag94,Lee06}. 
Even a single hole in a Mott insulator, i.e. a hole in an 
antiferromagnet in case of infinite Hubbard repulsion $U$, is substantially 
influenced by many-body effects \cite{Kane} because it's jump to a 
neighboring site disturbs antiferromagnetic arrangement of spins. 
Hence a thorough understanding of the dynamics of doped holes in Mott 
insulators has attracted a great deal of recent interest. 
The two major interactions relevant to the electrons in solids are 
electron-electron interactions (EEI) and electron-phonon interactions (EPI). 
The importance of the former at low doping is no doubt essential since
the Mott insulator is driven by strong Hubbard repulsion, while the latter was
considered to be largely irrelevant to superconductivity based on
the observations of a small isotope effect on the optimal $T_c$ 
\cite{isotope} and an absence of a phonon contribution to the resistivity
(for review see \cite{resistivity}). 

On the other hand,  there are now accumulating evidences that the EPI plays 
an important role in the physics of cuprates such as (i) an isotope effect 
on superfluid density $\rho_s$ and T$_{c}$ away from optimal doping 
\cite{Keller}, (ii)  neutron and Raman scattering \cite{Pint99,Raman,Raman1}
experiments showing strong phonon softening with both temperature and hole 
doping, indicating that EPI is strong \cite{Khal97,Gunnars2}.
Furthermore, the recent studies of cuprates by the angle resolved photoemission
spectroscopy (ARPES), which spectra are proportional to the LF 
(\ref{g}) \cite{Shen_03}, resulted in the discovery of the dispersion 
"kinks" at around 40-70meV measured from the Fermi energy, in the correct 
range of the relevant oxygen related phonons \cite{Lanzara01,Cuk2004,Dev2004}. 
These particular phonons - oxygen buckling and half-breathing modes are 
known to soften with doping \cite{Egami,Pint99} and with temperature
\cite{Cuk2004,Dev2004,Egami,Pint99,Raman,Raman1} indicating strong coupling. 
The quick change of the velocity can be predicted by any interaction of
a quasiparticle with a bosonic mode, either with a phonon
\cite{Cuk2004,Dev2004} or with a collective magnetic resonance mode
\cite{ChuNor,EschNor,EschNor1}.
However, the recently discovered ``universality'' of the kink energy for 
LSCO over the entire doping range \cite{Zhou03} casts doubts on the validity 
of the latter scenario as the energy scale of the magnetic excitation changes 
strongly with doping. 

Besides, measured in undoped high $T_c$ materials ARPES revealed apparent 
contradiction between momentum dependence of the energy and linewidth of 
the QP peak. 
On the one hand the experimental energy dispersion of the broad peak in 
many underdoped compounds \cite{ZX95,KyleDop} obeys the theoretical 
predictions \cite{Xiang_96,Kyu_Fer}, whereas the experimental peak width 
is comparable with the bandwidth and orders of magnitude larger than that 
obtained from theory of Mott insulator \cite{Hole01}.
Early attempts to interpret this anomalously short lifetime of a hole
by an interaction with additional nonmagnetic bosonic excitations, 
e.g.\ phonons \cite{Pothuizen97}, faced generic question: 
is it possible that interaction with media leaves the energy dispersion 
absolutely unrenormalized, while, induces a decay which inverse life-time is
comparable or even larger than the QP energy dispersion?
A possibility of an extrinsic origin of this width can be ruled out  
since the doping induces further disorder, while a sharper peak is observed 
in the overdoped region. 

In order to understand whether phonons can be responsible for peculiar
shape of the ARPES in the undoped cuprates, the LF of an interacting 
with phonons hole in Mott insulator was studied by DMC-SO \cite{tJpho}. 
The case of the LF of a single hole corresponds to the ARPES in an undoped 
compound. 
For a system with large Hubbard repulsion $U$, when $U$ is much larger than 
the typical bandwidth $W$ of noninteracting QP, the problem 
reduces to the $t$-$J$ model \cite{ChaSpaOles77,GJRice87,ManoRev94,Izymov97}
\begin{equation}
\hat{H}_{\mbox{\scriptsize t-J}} = - t \sum_{\langle ij \rangle s} 
                  c_{is}^{\dagger} c_{js}
                + J \sum_{\langle ij \rangle} 
\left(
{\bf S}_i {\bf S}_j - n_i n_j / 4
\right) \; . 
\label{tJ}
\end{equation}
Here $c_{j\sigma}$ is projected (to avoid double occupancy) 
fermion annihilation operator, $n_i$ ($< 2$) is the occupation number,
${\bf S}_i$ is spin 1/2 operator, $J$ is an exchange integral, and 
$\langle ij \rangle$ denotes nearest-neighbor sites in two dimensional 
square lattice. 
Different theoretical approaches revealed \cite{ManoRev94,Assad00,Hole01} 
basic properties of the LF. 
The LF has a sharp peak in the low energy part of the spectrum which 
disperses with a bandwidth $W_{J/t} \sim 2J$ and, therefore, the large
QP width in experiment can not be explained. 
More complicated $tt't''$-$J$ model takes into account hoppings to the second 
$t'$ and third $t''$ nearest neighbors and, hence, dispersion of the 
hole changes 
\cite{Chernyshov96_1,Chernyshov96_2,Tohyama_00,Xiang_96,Kyu_Fer,Shen_03}. 
However, for parameters, which are necessary for description of
dispersion in realistic high $T_c$ superconductors \cite{ZX95,Xiang_96}, 
peak in the low energy part remains sharp and well defined for all 
momenta \cite{Zaanen_95}.

After expressing spin operators in terms of Holstein-Primakoff spin wave
operators and diagonalizing the spin part of Hamiltonian (\ref{tJ}) by 
Fourier and Bogoliubov transformations \cite{SVR88,Kane,Liu_91,Liu_92}, 
$tt't''$-$J$ Hamiltonian is reduced to the boson-holon model, where  
hole (annihilation operator is $h_{\bf k}$) with dispersion 
%\begin{equation}
$\varepsilon({\bf k}) = 4t'\cos(k_x) \cos(k_y) +
              2t''[\cos(2k_x)+\cos(2k_y)] $
%\label{hole_disp}
%\end{equation} 
propagates in the magnon (annihilation operator is $\alpha_{\bf k}$) bath
\begin{equation}
\hat{H}_{\mbox{\scriptsize t-J}}^{0} =
\sum_{\bf k} \varepsilon({\bf k}) h_{\bf k}^{\dagger} h_{\bf k} 
+
\sum_{\bf k} \omega_{\bf k} \alpha_{\bf k}^{\dagger} \alpha_{\bf k}
\label{h0}
\end{equation}
with magnon dispersion $\omega_{\bf k}=2J\sqrt{1-\gamma_{\bf k}^2}$, where
$\gamma_{\bf k}=(\cos k_x + \cos k_y) / 2$. The hole is scattered by magnons
as described by 
\begin{equation}
\hat{H}_{\mbox{\scriptsize t-J}}^{\mbox{\scriptsize h-m}} =
N^{-1/2} \sum_{\bf k , q} M_{\bf k , q} 
\left[ h_{\bf k}^{\dagger} h_{\bf k-q} \alpha_{\bf k} + h.c.
\right] 
\label{h-m}
\end{equation}  
with the scattering vertex $M_{\bf k , q}$. 
Parameters $t$, $t'$ and $t''$ are hopping amplitudes to the first, 
second and third near neighbors, respectively. 
If hopping integrals $t'$ and $t''$ are set to zero and bare hole has no
dispersion, the problem (\ref{h0}-\ref{h-m}) corresponds to $t$-$J$ model.  

Short range interaction of a hole with dispersionless optical phonons 
%\begin{equation}
$\hat{H}^{\mbox{\scriptsize e-ph}} = 
\Omega_0 \sum_{\bf k} b_{\bf k}^{\dagger} b_{\bf k} $
%\label{phonons}
%\end{equation}
of the frequency $\Omega_0$ is introduced by Holstein Hamiltonian
\begin{equation}
\hat{H}^{\mbox{\scriptsize e-ph}} = 
N^{-1/2}  \sum_{\bf k , q} \frac{\sigma}{\sqrt{2M\Omega_0}} 
\left[ h_{\bf k}^{\dagger} h_{\bf k-q} b_{\bf q} + h.c.
\right] \; ,
\label{e-ph}
\end{equation}
where $\sigma$ is the momentum and isotope independent coupling 
constant, $M$ is the mass of the vibrating lattice ions, and 
$\Omega_0$ is the frequency of dispersionless phonon. 
The coefficient in front of square brackets is the standard 
Holstein interaction constant $\gamma=\sigma/\sqrt(2M\Omega_0)$.   
In the following we characterize strength of EPI in terms of dimensionless 
coupling constant $\lambda=\gamma^2/4t\Omega_0$. Note, if interaction 
with magnetic subsystem (\ref{h-m}) is neglected and hole dispersion
$\varepsilon({\bf k})$ is chosen in the form 
%\begin{equation}
$\varepsilon({\bf k}) = 2t[\cos(k_x)+\cos(k_y)]$,
%\label{holstein_disp}
%\end{equation} 
the problem (\ref{h0}), (\ref{e-ph}) corresponds to standard Holstein model 
where hole with near neighbor hopping amplitude $t$ interacts with 
dispersionless phonons. 

We consider the evolution of ARPES of a single hole in $t$-$J$-Holstein model 
(\ref{h0})-(\ref{e-ph}) from the weak to the strong coupling regime and 
dispersion of the LF in the strong coupling regime in Sect.\
\ref{sec_6_subsec_1}. 
It occurs that properties of the LF in the strong coupling regime of the EPI 
explain the puzzle of broad lineshape in ARPES in underdoped high $T_c$
superconductors.  
Therefore, in order to suggest a crucial test for the mechanism of 
phonon-induced broadening, we present calculations of 
the effect of the isotope substitution on the ARPES in Sect.\ 
\ref{sec_6_subsec_2}.

\subsection{Spectral Function of a Hole Interacting with Phonons  
in the $t$-$J$ Model: Self-Trapping and Momentum Dependence}
\label{sec_6_subsec_1}

Previously, the LF of $t$-$J$-Holstein model was studied by 
exact diagonalization 
method on small clusters \cite{Bauml_98} and in the non-crossing approximation
(NCA)\footnote{NCA is equivalent to self-consistent Born approximation 
(SCBA)} for both phonons and magnons \cite{Ramsak_92,Kyung_96}.
However, the small system size in exact diagonalization method implies a 
discrete spectrum and, therefore, the problem of lineshape could not be 
addressed. 
The latter method omits the FDs with mutual crossing of phonon propagators 
and, hence, is an invalid approximation for phonons in strong and 
intermediate couplings of EPI. 
This statement was demonstrated by DMC, which can sum all FDs for Holstein 
model  both exactly and in the NCA \cite{tJpho}. 
Exact results and those of NCA are in good agreement for small values 
$\lambda \le 0.4$ and drastically different for $\lambda>1$. 
For example,
for $\Omega_0/t=0.1$ exact result shows a sharp crossover to strong
coupling regime for $\lambda > \lambda_{H}^{c} \approx 1.2$ whereas
NCA result does not undergo such crossover even at $\lambda=100$. 
On the other hand, NCA is valid for interaction of a hole with magnons 
since spin S=1/2 can not flip more than once and number of magnons in the 
polaronic cloud can not be large. Note that the $t$-$J$-Holstein model is 
reduced 
to problem of polaron which interacts with several bosonic fields 
(\ref{3})-(\ref{4}).

DMC expansion in \cite{tJpho} takes into account mutual crossing of 
phonon propagators and, in the framework of partial NCA, neglects mutual 
crossing of magnon propagators, to avoid sign problem.
NCA for magnons is justified for $J/t \le 0.4$ by good agreement of results of 
NCA and exact diagonalization on small clusters
\cite{SVR88,Kane,Marsiglio91,MartHor_91,Liu_92}. 
Recently results of exact diagonalization were compared in the limit of small 
EPI for $t$-$J$-Holstein model, boson-holon model (\ref{h0}-\ref{e-ph}) without 
NCA, and  boson-holon model with NCA  \cite{RoGu_SCBA}. 
Although agreement is not so good as for pure $t$-$J$ model, it was concluded 
that NCA for magnons is still good enough to suggest that one can use NCA for 
a qualitative description of the $t$-$J$-Holstein model.  

%%%%%%%%%%%%%%%%%%%%%%%%%%%%%%%%%%%%%%%%%
\begin{figure}[htb]
\hspace{0.0 cm}  \vspace {-1.0 cm}
\includegraphics{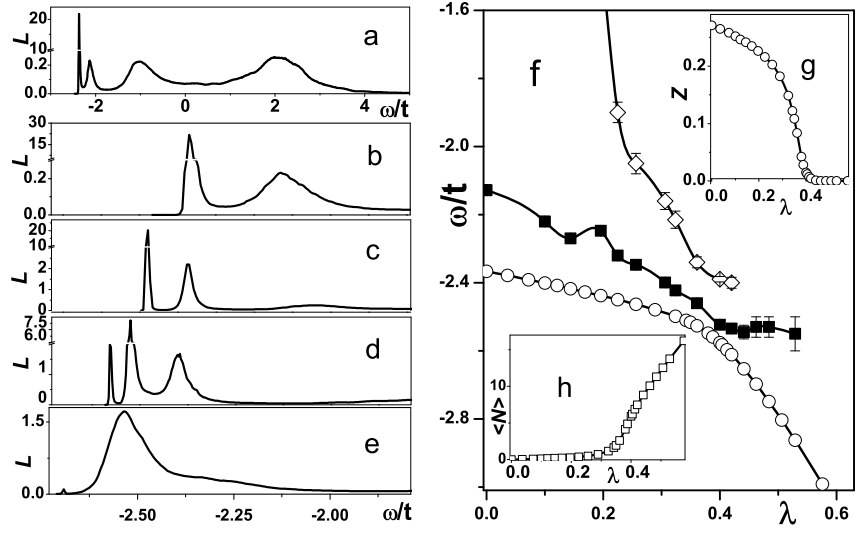}
\caption{\label{fig:vartj_1} (a) The LF of a hole in the ground state 
${\bf k}=(\pi/2,\pi,2)$ at $J/t=0.3$ and $\lambda=0$. 
Low energy part of the LF of a hole in the 
ground state ${\bf k}=(\pi/2,\pi,2)$ 
at $J/t=0.3$: (b) $\lambda=0$; (c) $\lambda=0.3$;  
(d) $\lambda=0.4$; (e) $\lambda=0.46$.
Dependence on coupling strength $\lambda$ at $J/t=0.3$:
(f) energies of lowest LF resonances;
(g) $Z$-factor of lowest peak; 
(h) average number of phonons 
$\langle N \rangle$.
} 
\end{figure}
%%%%%%%%%%%%%%%%%%%%%%%%%%%%%%%%%%%%%%%%%

\begin{figure}[htb]
\hspace{0.0 cm}  \vspace {-0.5 cm}
\includegraphics{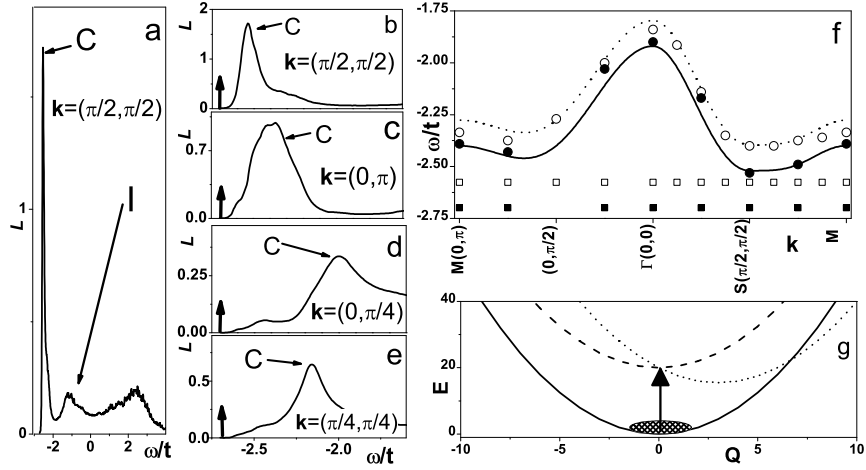}
\caption{\label{fig:vartj_2} 
The LF of a hole at $J/t=0.3$ and $\lambda=0.46$:
(a) full energy range for ${\bf k}=(\pi / 2, \pi / 2)$;
(b--e) low energy part for different momenta.
Slanted arrows show broad peaks which can be interpreted in ARPES
spectra  as coherent (C) and incoherent (I) part.
Vertical arrows in panels (b)--(e)
indicate position of ``invisible'' lowest resonance.
(f) Dispersion of resonances energies at $J/t=0.3$:
broad resonance (filled circles) and lowest polaron pole
(filled squares) at $\lambda=0.46$; broad resonance
(open circles) and lowest polaron pole (open squares)
at $\lambda=0.4$.
The solid curves are dispersions (\ref{Marsig}) of a hole in pure
$t$-$J$ model at $J/t=0.3$ ($W_{J/t=0.3}=0.6$):
$\varepsilon_{min}=-2.396$ ($\varepsilon_{min}=-2.52$) for dotted
(solid) line.
Panel (g) shows ground state potential $Q^2/2$ (solid line), 
excited state potential without relaxation $D+Q^2/2$ (dashed line),
and relaxed excited state potential $D+(Q-\lambda)^2/2-\lambda^2/2$ 
(dotted line). }
\end{figure}

Figures~\ref{fig:vartj_1}a-e show 
low energy part of LF in the ground state at  
${\bf k}=(\pi / 2, \pi / 2)$ in the weak, intermediate, and strong coupling
regimes of interaction with phonons. 
Dependence on the coupling constant of energies of resonances 
(Fig.~\ref{fig:vartj_1}f), $Z^{{\bf k}=(\pi/2,\pi/2)}$-factor of lowest peak 
(Fig.~\ref{fig:vartj_1}g), and average number of phonons in the polaronic 
cloud $\langle N \rangle$ (Fig.~\ref{fig:vartj_1}h) demonstrates a picture 
which is typical for ST (see \cite{R82,RP02} and Sect.\ \ref{section_4}).
Two states cross and hybridize in the vicinity of critical coupling
constant  $\lambda_{\mbox{\scriptsize t-J}}^{c} \approx 0.38$,
$Z^{{\bf k}=(\pi/2,\pi/2)}$-factor of lowest resonance sharply drops and
average number of phonons in polaronic cloud quickly rises. 
According to the general understanding of the ST phenomenon, above 
the critical couplings 
$\lambda>\lambda_{\mbox{\scriptsize t-J}}^{c}$ one expects that the lowest 
state is dispersionless while the upper one has small effective mass. 
This assumption is supported by the momentum dependence of the LF in the 
strong coupling regime (Fig.~\ref{fig:vartj_2}a-e).  
Dispersion of upper broad shake-off Franck-Condon peak nearly perfectly obeys 
relation
\begin{equation}
\varepsilon_{\bf k} = \varepsilon_{min} +
W_{J/t} / 5
\{
[\cos k_x + \cos k_y]^2  +
[\cos (k_x+k_y) + \cos (k_x-k_y)]^2/4 
\},
\label{Marsig}
\end{equation}     
which describes dispersion of the pure $t$-$J$ model in the broad range of
exchange constant $0.1<J/t<0.9$ \cite{Marsiglio91} 
(Fig.~\ref{fig:vartj_2}f). Note that this property of the shake-off peak 
is general for the whole strong coupling regime 
(Fig.~\ref{fig:vartj_2}f).

%%%%%%%%%%%%%%%%%%%%%%%%%%%%%%%%%%%%%%%%%
%%%%%%%%%%%%%%%%%%%%%%%%%%%%%%%%%%%%%%%%%

Momentum dependence of the shake-off peak, reproducing that of the 
free particle, is the direct consequence of the adiabatic regime. 
Actually, phonon frequency $\Omega_0$ is much smaller than the coherent
bandwidth $2J$ of the $t$-$J$ model, giving the adiabatic ratio
$\Omega / 2J = 1/6 \ll 1$. 
Besides, as experience with the OC of the Fr\"{o}hlich polaron (Sect.\
\ref{sec_3_subsec_2}) shows, there is one more important parameter in
the strong coupling limit. 
Namely, the ratio between measurement process time $\tau_{mp}=\hbar / \Delta E$
where $\Delta E$ is the energy separation of shake-off hump from the 
ground state pole, and that of characteristic lattice time 
$\tau \approx 1 / \Omega_0$ is much less than unity. 
Hence, fast photoemission probe sees the ions frozen in one of possible 
configurations \cite{Misko97}. 
The LF in the FC limit is a sum of transitions between a lower 
$E_{\mbox{\scriptsize low}}(Q)$ and an upper 
$E_{\mbox{\scriptsize up}}(Q)$ sheets of adiabatic potential,
weighted by the adiabatic wave function of the lower sheet 
$\mid \psi_{\mbox{\scriptsize low}}(Q) \mid^2$ \cite{RoGu2005}. 
If EPI is absent both in initial $E_{\mbox{\scriptsize low}}(Q)= Q^2 /2$ and
final $E_{\mbox{\scriptsize up}}(Q)= {\cal D} + Q^2 /2$ states, 
the LF is peaked at the energy ${\cal D}$. Then, if there is EPI 
$\Delta E_{\mbox{\scriptsize up}}(Q)= - \lambda Q$ only in the final state,
i.e. when hole is removed from the Mott insulator, 
the upper sheet of adiabatic potential 
$E_{\mbox{\scriptsize up}}(Q)= {\cal D} -\lambda^2/2 + (Q-\lambda)^2 /2$
has the same energy ${\cal D}$ at $Q=0$. Since the probability 
function $\mid \psi_{\mbox{\scriptsize low}}(Q) \mid^2$ has maximum 
at $Q=0$, the peak of the LF broadens but it's energy does not shift
\cite{RoGu2005} (Fig. \ref{fig:vartj_2}g).   
 
Behavior of the LF is the same as observed in the ARPES of undoped cuprates. 
The LF consists of a broad peak and a high energy incoherent continuum 
(see Fig.~\ref{fig:vartj_2}a). 
Besides, dispersion of the broad peak ``c'' in Figs.~\ref{fig:vartj_2} 
reproduces that of sharp peak in pure $t$-$J$ model (Fig.~\ref{fig:vartj_2}b-f). 
The lowest dispersionless peak, corresponding to small radius polaron, has 
very small weight and, hence, can not be seen in experiment. 
On the other hand, according to experiment, momentum dependence of spectral 
weight $Z^{({\bf k})'}$ of broad resonance exactly  reproduces dispersion 
of $Z^{({\bf k})}$-factor of pure $t$-$J$ model.
The reason for such perfect mapping is that in adiabatic case 
$\Omega_0 /2J \ll 1$ all weight of the sharp resonance in $t$-$J$ model without
EPI is transformed at strong EPI into the broad peak.   
This picture implies that the chemical potential in the heavily
underdoped cuprates is not connected with the broad resonance but 
pinned to the real quasiparticle pole with small $Z$-factor. 
This conclusion was recently confirmed experimentally \cite{KyleDop}.  

Comparing the critical EPI for a hole in the $t$-$J$-Holstein model
(\ref{h0}-\ref{e-ph}) $\lambda_{\mbox{\scriptsize t-J}}^{c} \approx 0.38$ 
and that for Holstein model $\lambda_{H}^{c} \approx 1.2$ with the same 
value of hopping $t$, we conclude that \textit{spin-hole interaction 
accelerates transition into the strong coupling regime}. 
The reason for enhancement of the role of EPI is found in 
\cite{RoGu_SCBA}. 
Comparison of the EPI driven renormalization of the effective mass in 
$t$-$J$-Holstein and Holstein model shows that large effective mass in the 
$t$-$J$ model is responsible for this effect. 
The enhancement of the role of EPI by EEI takes place at least for a single 
hole at the bottom of the $t$-$J$ band. 
Had the comparison been made with half-filled model, the result would have 
been smaller enhancement or no enhancement at all \cite{Sangio_Gunn_06}.
On the other hand, coupling constant of half-breathing phonon is increased
by correlations \cite{RoGuPRB_04}. Finally, we conclude that effect 
of enhancement of the effective EPI by EEI is not unambiguous and depends on 
details of interaction and filling. 
However, this effect is present for small filling in the $t$-$J$-Holstein model.

\subsection{Isotope Effect on ARPES in Underdoped High-Temperature
Superconductors}
\label{sec_6_subsec_2}

The magnetic resonance mode and the phonon modes are the two major candidates
to explain the ``kink'' structure of the electron energy dispersion 
around 40-70 
meV below the Fermi energy, and the isotope effect (IE) on ARPES should be 
the smoking-gun experiment to distinguish between these two. 
Gweon et al. \cite{Gweon_04} performed the ARPES experiment on 
O$^{18}$-replaced Bi2212 at optimal doping and found an appreciable IE, 
which however can not be explained within the conventional weak-coupling 
Migdal-Eliashberg theory. 
Namely the change of the spectral function due to O$^{18}$-replacement has 
been observed at higher energy region beyond the phonon energy ($\sim 60$meV). 
This is in sharp contrast to the weak coupling theory prediction, i.e., 
the IE should occur only near the phonon energy. 
Hence the IE in optimal Bi2212 remains still a puzzle. 
On the other hand, the ARPES in undoped materials, as described
in Sect.\ \ref{sec_6_subsec_1}, has recently been understood 
in terms of the small polaron formation  \cite{tJpho,RoGuZX,RoGu2005}.
Therefore, it is essential to compare experiment in undoped systems with 
presented in this Sect. DMC-SO data, where theory can offer quantitative 
results.  

In addition to high-$T_c$ problem, strong EPI mechanism of ARPES
spectra broadening was considered as one of alternative scenarios
for diatomic molecules \cite{Zawatzky_89},
colossal magnetoresistive manganites \cite{Dessau_98},
quasi-one-di\-men\-si\-o\-nal Peierls conductors \cite{Perf_01,Perf_02},
and Verwey magnetites \cite{Schrupp_05}.
Therefore, exact analysis of the IE on ARPES at strong EPI is of
general interest for conclusive experiments in a broad variety 
of compound classes.      

Dimensionless coupling constant $\lambda=\gamma^2/4t\Omega$ in (\ref{e-ph}) 
is an invariant quantity for the simplest case of IE. 
Indeed, assuming natural relation $\Omega \sim 1/\sqrt{M}$ 
between phonon frequency and mass, we find that $\lambda$ does not depend 
on the isotope factor 
$\kappa_{\mbox{\scriptsize iso}} = \Omega/\Omega_0 = \sqrt{M_0/M}$, which 
is defined as the ratio of phonon frequency in isotope substituted 
($\Omega$) and normal ($\Omega_0$) systems.
We chose adopted parameters of the $tt't''$-$J$ model which reproduce 
the experimental dispersion of ARPES \cite{Xiang_96}: 
$J/t=0.4$, $t'/t=-0.34$, and $t''/t=0.23$ . 
The frequency of the relevant phonon \cite{Shen_03} is set to 
$\Omega_0/t=0.2$ and the isotope factor 
$\kappa_{\mbox{\scriptsize iso}}=\sqrt{16/18}$
corresponds to substitution of O$^{18}$ isotope for O$^{16}$. 

To sweep aside any doubts of possible instabilities of analytic 
continuation, we calculate the LF for normal compound 
($\kappa_{\mbox{\scriptsize nor}}=1$), isotope substituted 
($\kappa_{\mbox{\scriptsize iso}}=\sqrt{16/18}$) 
and ``anti-isotope'' substituted 
($\kappa_{\mbox{\scriptsize ant}}=\sqrt{18/16}$) compounds. 
Monotonic dependence of LF on $\kappa$ ensures stability of analytic 
continuation and gives possibility to evaluate the error-bars of a quantity
${\cal A}$ using quantities 
${\cal A}_{\mbox{\scriptsize iso}}-{\cal A}_{\mbox{\scriptsize nor}}$,       
${\cal A}_{\mbox{\scriptsize nor}}-{\cal A}_{\mbox{\scriptsize ant}}$, and 
$({\cal A}_{\mbox{\scriptsize iso}}-{\cal A}_{\mbox{\scriptsize ant}})/2$.

%%%%%%%%%%%%%%%%%%%%%%%%%%%%%%%%%%%%%%%%%
%\begin{figure}[htb]
%\hspace{0.0 cm}  \vspace {-1.0 cm}
%\includegraphics{isot_1.eps}
%\caption{\label{fig:isot_1} Panels (a--c): hole 
%LFs at ${\bf k}=(\pi/2,\pi/2)$ (solid line) and ${\bf k}=(\pi,0)$ 
%(dashed line) for different couplings.
%Panels (d--e):  low energy part of LF for $\lambda=0.62$
%(solid line), $\lambda=0.69$ (dashed line), and $\lambda=0.75$ 
%(dotted line) at nodal (d) and antinodal (e) points.
%} 
%\end{figure}
%%%%%%%%%%%%%%%%%%%%%%%%%%%%%%%%%%%%%%%%%

%%%%%%%%%%%%%%%%%%%%%%%%%%%%%%%%%%%%%%%%%
\begin{figure}[htb]
\hspace{0.0 cm}  \vspace {-1.0 cm}
\includegraphics{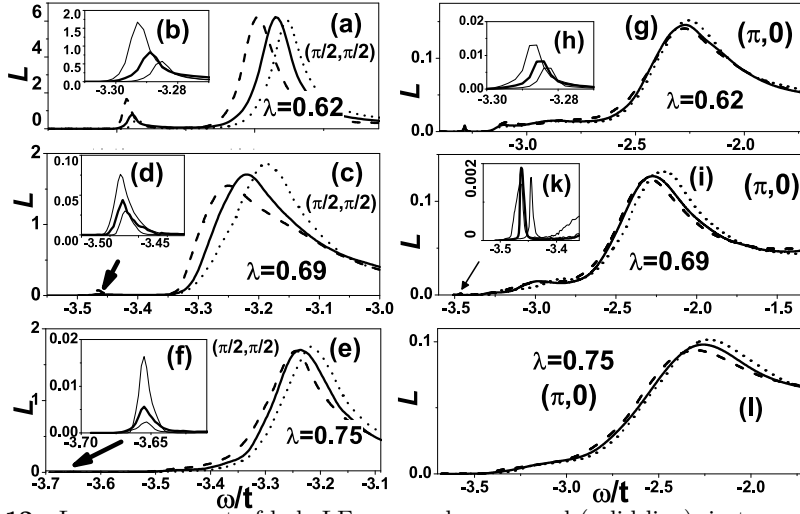}
\caption{\label{fig:isot_2} Low energy part of hole LFs:
normal compound (solid line), isotope substituted compound (dotted line) and
``antiisotope'' substituted compound (dashed line). 
LFs at different couplings in the nodal (a, c, e) and 
antinodal (g, i, l) points. 
Insets (b, d, f, h, k) show low energy peak of real QP. 
} 
\end{figure}
%%%%%%%%%%%%%%%%%%%%%%%%%%%%%%%%%%%%%%%%%

Since LF is sensitive to strengths of EPI only for low frequencies 
\cite{Isotope}, we concentrate on the low energy part of the spectrum. 
Figure \ref{fig:isot_2} shows IE on the hole LF 
for different couplings in nodal and antinodal points, respectively. 
The general trend is a shift of all spectral features to larger 
energies with increase of the isotope mass ($\kappa<1$). 
One can also note that the shift of broad FCP is much larger than that 
of narrow real-QP peak. 
Moreover, for large couplings $\lambda$ the shift of QP energy 
approaches zero and only decrease of QP spectral weight $Z$ 
is observed for larger isotope mass. 
On the other hand, the shift of FCP is not suppressed for larger 
couplings. Except for the LF in nodal point at $\lambda=0.62$ 
(Fig.~\ref{fig:isot_2}a, b), where LF still has significant weight of QP 
$\delta$-functional peak, there is one more notable feature of the IE.
With increase of the isotope mass the height of FCP increases.
Taking into account the conservation law for LF 
$\int_{-\infty}^{+\infty}L_{\bf k}(\omega)=1$ and insensitivity 
of high energy part of LF to EPI strength \cite{Isotope}, 
the narrowing of the FCP for larger isotope mass can be concluded.
To understand the trends of the IE in the strong coupling regime
we analyze the exactly solvable independent oscillators model
(IOM) \cite{Mahan}. The LF in IOM is the Poisson distribution
\begin{equation}
L(\omega) = \exp[-\xi_0/\kappa] \sum_{l=0}^{\infty} 
\frac{[\xi_0/\kappa]^l}{l!} {\cal G}_{\kappa,l}(\omega) \; ,
\label{lsf}
\end{equation}
where $\xi_0=\gamma^2_0/\Omega_0^2=4t\lambda/\Omega_0$ is 
dimensionless coupling constant for normal system and 
${\cal G}_{\kappa,l}(\omega) = \delta[\omega+4t\lambda-\Omega_0\kappa l]$
is the $\delta$-function. 
The properties of the Poisson distribution quantitatively explain 
many features of the IE on LF\footnote{Cautions should be 
made about approximate form of EPI (\ref{e-ph}). 
Strictly speaking, actual momentum dependence of the interaction
constant $\sigma$ \cite{RoGu_04,Ishihara_04} can slightly change the
obtained differences between nodal and antinodal points though
the general trends have to be left intact because ST is caused 
solely by the short range part of EPI \cite{R82}.}.

%%%%%%%%%%%%%%%%%%%%%%%%%%%%%%%%%%%%%%%%%
\begin{figure}[tbh]
\hspace{-0.3 cm}  \vspace {-0.5 cm}
\includegraphics{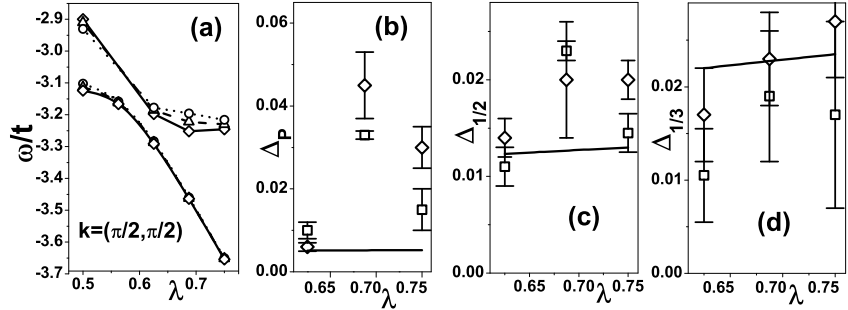}
\caption{\label{fig:isot_3} 
(a) Energies of ground state and broad peaks
for normal (triangles), isotope substituted
(circles) and ``antiisotope'' substituted (diamonds) compounds.
Comparison of IOM estimates (lines) with 
DMC data in the nodal (squares) and antinodal (diamonds) points:
(b) shift of the FCP top, (c) FCP leading edge at $1/2$ of height,
and (d) FCP leading edge at $1/3$ of height.}
\end{figure}
%%%%%%%%%%%%%%%%%%%%%%%%%%%%%%%%%%%%%%%%%

The energy $\omega_{\mbox{\scriptsize QP}}=-4t\lambda$ of the zero-phonon 
line $l=0$ in (\ref{lsf}) depends only on isotope independent quantities
which explains very weak isotope dependence of QP peak energy in insets of  
Fig.~\ref{fig:isot_2}. 
Besides, change of the zero-phonon line weight 
$Z^{(0)}$ obeys relation 
$Z^{(0)}_{\mbox{\scriptsize iso}}/Z^{(0)}_{\mbox{\scriptsize nor}} =
\exp\left[ -\xi_0(1-\kappa)/\kappa\right]$ in IOM. 
These IOM estimates agree with DMC data within 15\% in the nodal point and 
within 25\% in the antinodal one.  
IE on FCP in the strong coupling regime follows from the properties 
of zero
$M_0=\int_{-\infty}^{+\infty} L(\omega) d \! \omega =1$,
first 
$M_1=\int_{-\infty}^{+\infty} \omega L(\omega) d \! \omega =0$, and second
$M_2=\int_{-\infty}^{+\infty} \omega^2 L(\omega) d \! \omega 
= \kappa \xi_0 \Omega_0^2$ moments of shifted Poisson distribution 
(\ref{lsf}). 
Moments $M_0$ and $M_2$ establish relation  
${\cal D} = h^{\mbox{\scriptsize FCP}}_{\mbox{\scriptsize iso}} /
h^{\mbox{\scriptsize FCP}}_{\mbox{\scriptsize nor}} = 1/\sqrt{\kappa}
\approx 1.03$ between heights of FCP in normal and substituted compounds.
DMC data in the antinodal point perfectly agree with the above estimate 
for all couplings. 
This is consistent with the idea that the anti-nodal region remains 
in the strong coupling regime even though the nodal region is in the 
crossover region.
In the nodal point DMC data well agree with IOM estimate for 
$\lambda=0.75$ 
(${\cal D} \approx 1.025$) whereas at $\lambda=0.69$ and $\lambda=0.62$ 
influence of the ST point leads to anomalous values of 
${\cal D}$: ${\cal D} \approx 1.07$ and ${\cal D} \approx 0.98$, respectively.
Shift of the low energy edge at half maximum $\Delta_{1/2}$ must be  
proportional to change of the root square of second moment 
$\Delta_{\sqrt{M_2}} = \sqrt{\xi_0} \Omega_0 [1-\sqrt{\kappa}]$. 
As we found in numeric simulations of (\ref{lsf}) with Gaussian 
functions\footnote{ Results are almost independent on the parameter
$\eta$ of the Gaussian distribution
${\cal G}_{\kappa,l}(\omega) 
= 1/(\eta\sqrt{2\pi}) \exp(-[\omega+4t\lambda-\Omega_0\kappa l]/(2\eta^2))$ 
in the range $[0.12,0.2]$. } 
${\cal G}_{\kappa,l}(\omega)$, 
relation $\Delta_{1/2} \approx \Delta_{\sqrt{M_2}} / 2$ is accurate to 10\% 
for $0.62<\lambda<0.75$. 
Also, simulations show that the shift of the edge at one third of 
maximum $\Delta_{1/3}$ obeys relation 
$\Delta_{1/3} \approx \Delta_{\sqrt{M_2}}$. 
DMC data with IOM estimates are in good agreement for strong EPI 
$\lambda=0.75$ (Fig.~\ref{fig:isot_3}). However, shift of the FCP top 
$\Delta_p$ and $\Delta_{1/2}$ are considerably enhanced in the 
self-trapping (ST) transition region. 
The physical reason for enhancement of IE in this region
is a general property regardless of the QP dispersion, range of EPI, etc.  
The influence of nonadiabatic matrix element, mixing excited and ground 
states, on the energies of resonances essentially depends on the 
phonon frequency. 
While in the adiabatic approximation ST transition is sudden and 
nonanalytic in $\lambda$ \cite{R82}, nonadiabatic matrix elements turn 
it to smooth crossover \cite{Gerlach}. 
Thus, as illustrated in Fig.~\ref{fig:isot_3}a, the smaller the frequency 
the sharper the kink in the dependence of excited state energy on the 
interaction constant

In the undoped case the present results can be directly compared with the 
experiments.
It is found that the IE on the ARPES lineshape of a single hole
is anomalously enhanced in the intermediate coupling regime while can be 
described by the simple independent oscillators model in the strong 
coupling regime. 
The shift of FCP top and change of the FCP height are relevant quantities 
to pursue experimentally in the intermediate coupling regime since IE 
on these characteristics is enhanced near the self trapping point. 
In contrast, shift of the leading edge of the broad peak is the relevant 
quantity in the strong coupling regime since this value increases with 
coupling as $\sqrt{\lambda}$. 
These conclusions, depending on the fact whether self trapping phenomenon 
is encountered in specific case, can be applied fully or partially to another 
compounds with strong EPI \cite{Dessau_98,Perf_01,Perf_02,Schrupp_05}.

\subsection{Conclusions and Perspectives}
\label{sec_6_subsec_3}

In this article, we have focused mainly on the polaron problem in strongly 
correlated systems. 
This offers an approach from the limit of low carrier concentration doped into 
the (Mott) insulator, which is complementary to the conventional 
Eliashberg-Migdal approach for the EPI in metals. 
In the latter case, we have the Fermi energy $\varepsilon_F$ as a relevant 
energy scale, which is usually much larger than the phonon frequency 
$\Omega_0$. 
In this case, the adiabatic Migdal approximation is valid and the vertex 
corrections, which correspond to the multi-phonon cloud and are essential to 
the self-trapping phenomenon, are suppressed by the ratio 
$\Omega_0/\varepsilon_F$. 
Therefore an important issue is the crossover from the strong coupling 
polaronic picture to the weak coupling Eliashberg-Migdal picture. 
This occurs as one increases the carrier doping into the insulator. 
As is observed by ARPES experiments in high temperature superconductors, 
the polaronic states continue to survive even at finite doping 
\cite{KyleDop}. 
This suggests a novel polaronic metallic state in underdoped cuprates, which 
is common also in CMR manganites \cite{Manella_05} 
and is most probably universal in transition metal oxides. 
In the optimal and overdoped region, the Eliashberg-Migdal picture becomes 
appropriate \cite{Cuk2004,Dev2004}, 
but still a nontrivial feature of the EPI is its strong momentum dependence 
leading to the dichotomy between the nodal and anti-nodal regions. 
It is an interesting observation that the highest superconducting transition 
temperature is attained at the crossover region between the two pictures 
above, which suggests that both the itinerancy and strong coupling to the 
phonons are essential to the quantum coherence. 
It should be noted that this crossover occurs in a nontrivial way also in the 
momentum space, i.e., the nodal and anti-nodal 
regions behave quite differently as discussed in Sect.\ 
\ref{sec_6_subsec_2}. 
However, the relevance of the EPI to the high $T_c$ superconductivity is still 
left for future investigations.

We hope that this article convinces the readers the vital role of ARPES 
experiments and numerically exact solutions to the EPI problem, 
the combination of them offers a powerful tool for the momentum-energy 
resolved analysis of these rather complicated strongly correlated electronic 
systems. This will pave a new path to the deeper understanding of the 
many-body electronic systems. 

We thank Y. Toyozawa, Z. X. Shen, T. Cuk, T. Devereaux, J. Zaanen, 
S. Ishihara, A. Sakamoto, N. V. Prokof'ev, B. V. Svistunov, E. A. Burovski, 
J. T. Devreese, G. de Filippis, V. Cataudella, P. E. Kornilovitch,
O. Gunnarsson, N. M. Plakida, and K. A. Kikoin, for collaborations 
and discussions. 

%
%
% BibTeX users please use
% \bibliographystyle{}
% \bibliography{}
%
% Non-BibTeX users please follow the syntax
% the syntax of "referenc.tex" for your own citations
%%%%%%%%%%%%%%%%%%%%%%%% referenc.tex %%%%%%%%%%%%%%%%%%%%%%%%%%%%%%
% sample references
% "physics"
%
% Use this file as a template for your own input.
%
%%%%%%%%%%%%%%%%%%%%%%%% Springer-Verlag %%%%%%%%%%%%%%%%%%%%%%%%%%

%
% BibTeX users please use
% \bibliographystyle{}
% \bibliography{}

%%%%%%%%%%%%%%%%%%%%%%%%%%%%%%%%%%%%%%%%%%%%%%%%%%%%%%%%%%%%%%%%%%%%%%  }

%%%%%%%%%%%%%%%%%%%%%%%%%%%%%%%%%%%%%%%%%%%%%%%%%%%%%%%%%%%%%%%%%%%%%%

\end{document}